\documentclass[12pt]{iopart}
\usepackage{amsfonts}
\usepackage{amssymb}
\usepackage{epsfig}
\usepackage{graphicx}
\usepackage{subfigure}
\usepackage{mathrsfs}
\usepackage{mathrsfs}

\DeclareRobustCommand{\scrD}{\mathscr{D}}

\def\afrac#1/#2{\leavevmode\kern.1em
\raise.5ex\hbox{\the\scriptfont0 #1}\kern-.1em
/\kern-.15em\lower.25ex\hbox{\the\scriptfont0 #2}}

\def\safrac#1/#2{\leavevmode\kern.1em
\raise.25ex\hbox{\the\scriptscriptfont0 #1}\kern-.1em
/\kern-.15em\lower.25ex\hbox{\the\scriptscriptfont0 #2}}

\usepackage{citesort}

\def\beq{\begin{equation}}
\def\eeq{\end{equation}}
\def\beqa{\begin{eqnarray}}
\def\eeqa{\end{eqnarray}}

\def\half{{1\over2}}

\def\Hb{\mathbf{H}}
\def\Cb{\mathbb{C}}
\def\Rb{\mathbb{R}}

\DeclareRobustCommand{\scrD}{\mathscr{D}}

\def\t{\textstyle}

\font\ftfrak=eufm10 at 10 pt
\font\lfrak=eufm10 at 12 pt
\font\Lfrak=eufm10 at 12 pt
\newcommand{\fr}[1]{\mbox{\ftfrak #1}}
\newcommand{\lfr}[1]{\mbox{\lfrak #1}}
\newcommand{\Lfr}[1]{\mbox{\Lfrak #1}}

\newcommand\eq[1]{Eqn.~(\ref{#1})}

\def\spacer#1#2 {\rule[#1]{0in}{#2}}

\begin{document}

Published in J.\ Phys.\ A: Math,\ Theor.\ {\bf 45} (2012) 24003 

\title{Vector coherent state representations and their inner products}

\author{D.J. Rowe }

\address{Department of Physics, University of Toronto, Toronto, ON M5S 1A7, Canada%
}

\
\date{Feb, 2011}

\begin{abstract}
Several advances have extended the power and versatility of coherent state  theory  to the extent that it has become a vital tool in the representation theory of Lie groups and their Lie algebras.  
Representative applications are reviewed  
and some new developments are introduced. 
The examples given  are chosen to illustrate special features of the scalar and vector coherent state constructions and how they work in practical situations.  Comparisons are made with Mackey's theory of induced representations.
 For simplicity, we focus  on square integrable (discrete series) unitary representations although many of the techniques apply more generally,  with minor adjustment.
 \\ \\
\noindent {PACS numbers:} { 02.20.--a, 02.20.Sv, 21.60.Fw,  21.60.Ev}
\end{abstract}



\section{Introduction}

Coherent state theory is known to reveal  classical behaviour in quantum mechanics.  
Thus, it has been used extensively in studying the relationship between classical and quantum mechanics and in the development of  quantization techniques   
\cite{KlauderS85,Berezin75,AliE05,Gazeau09,RoweCQM}.
The applications of coherent state theory outlined in this paper facilitate the quantisation of an algebraic system by construction of the unitary representations of its spectrum generating algebra. 
 
Standard coherent state representations, which we  refer to as 
\emph{scalar coherent state} representations, were introduced by Bargmann 
\cite{Bargmann61}
and Segal \cite{Segal63} and defined more generally by Perelomov 
\cite{Perelomov72}, Onofri \cite{Onofri75}, and others.
They were subsequently extended to \emph{vector-valued coherent state} (VCS) representations  \cite{Rowe84,RoweRC84,RoweRG85}
and used widely in the construction of explicit
representations of many Lie algebras and Lie groups 
\cite{HechtLeBR87,RoweLeBH87,LeBlancR88,LeBlancR89,LeBlancR90}.
Early applications were reviewed in a book by Hecht \cite{Hecht87}.
A new class of representations was introduced for representations on functions of SO(3) \cite{RoweLeBR89,RoweVC89,RoweH95,TurnerRR06}
which were later used in the construction of  shift tensors \cite{RoweR95}, and for the computation of Clebsch-Gordan coefficients for reducing tensor product representations \cite{RoweR97,RoweB00,BahriRD04}.  VCS theory was, in fact, designed for the specific purpose of inducing the harmonic series of irreps 
 (irrreducible representations)   
 of the non-compact symplectic Lie algebra $\lfr{sp}(3,\Rb)$ from those of its maximal compact  \lfr{u}(3) sub\-algebra 
 \cite{Rowe84a,RoweRC84,RoweRG85,Rowe84}.
These irreps  are needed in applications of the microscopic nuclear collective model \cite{RosensteelR77,RosensteelR80,Rowe85}.

It has been shown  \cite{RoweR91} that  VCS theory is a physically intuitive theory of   induced representations  \cite{Mackey68}
with the advantage that the representations it induces are irreducible. 
It is also known \cite{BartlettRR02a,BartlettRR02b} that scalar and VCS representations relate closely to those of geometric quantization 
\cite{Souriau69,Kostant70a,Kostant70b}
and, in some respects, extend them.

In this review, we restrict consideration to applications of scalar and VCS theory to the construction of unitary irreps of Lie algebras.  Several examples are used to illustrate the different ways and the much larger variety of representations that can be induced by the extension to vector-valued wave functions.  The  construction of a VCS representation is  straightforward.  The more  challenging part is to determine an orthonormal basis for its Hilbert space and calculate the matrix elements  required for the application of spectrum generating algebras in quantum physics.  For this purpose, VCS theory relies on K-matrix theory \cite{Rowe84,RoweR91,Rowe95}. 
K-matrix theory 
  (distinct from K-theory as used in mathematics)  
provides practical procedures for determining the inner products of coherent state and VCS representations as shown, in the context in which it is used, in this review.
It is an essential component of VCS theory, although it can be used  more generally. 
It could  be used, for example, to compute  matrix elements of a unitary representation from those obtained by the partial-coherent-state methods of Deenen and Quesne \cite{DeenenQ84}.
{  K-matrix theory is developed in Sect.\ \ref{sect:KmatTheory}.

In approaching coherent-state representation theory as a theory of quantisation, it is useful to recognise that a  Schr\"odinger representation of state vectors in quantum mechanics by wave functions is, in fact, a coherent state representation.  This is explained in Section \ref{sect:conc}.

\section{Scalar coherent state representation}

Many definitions of coherent states and  coherent state representations have been given \cite{Klauder63a,Perelomov72,Gilmore72} and are described in several reviews \cite{KlauderS85,Perelomov85,ZhangFG90}.  
Following Perelomov  \cite{Perelomov85}, we start with a basic definition but quickly adjust it to one that is more useful.

\subsection{Basic coherent state representations}

Let $\hat T$ denote a unitary irrep  of a Lie group $G$ on a Hilbert space $\Hb$ with inner product of two vectors $|\psi\rangle$  and
$|\varphi\rangle$ denoted by $\langle \psi|\varphi\rangle$. 
Then, for any normalised   vector $|\phi\rangle\in \Hb$ (known as a \emph{fiducial  vector}),  a system of coherent state vectors for the irrep is defined as the set
\beq \mathfrak{M}_\phi := \{ |\phi(g)\rangle = \hat T(g) |\phi \rangle; g\in G\} .
\label{eq:1.CSvectors}\eeq
The  vectors in $\mathfrak{M}_\phi$ span the Hilbert space $\Hb$.  
Thus, in a coherent-state representation,
an arbitrary  vector $|\psi\rangle\in\Hb$ can be assigned a 
wave function, $\Psi$, defined on the group $G$ by the overlaps
\beq \Psi(g) := \langle \phi |\hat T(g) |\psi\rangle , \quad g\in G. \eeq
The space spanned by these wave functions, $\mathcal{H}$, carries a coherent-state irrep $\hat \Gamma$, that is isomorphic to $\hat T$ and defined by
\beq \hat\Gamma(g) \Psi(g') := \langle \phi| \hat T(g') \hat T(g) |\psi\rangle = \Psi(g'g),   \quad \forall \, g,g' \in G .  \label{eq:1.GammaG}\eeq
Thus, $\mathcal{H}$ is a Hilbert space  with  inner product inherited from the map  $\Hb \to \mathcal{H} ; |\psi\rangle \mapsto \Psi$. 

When 
  $G$ is compact, and for some (e.g., discrete series) representations when $G$ is non-compact, the inner product for $\mathcal{H}$ is
is defined by the so-called \emph{resolution of the identity} operator
\beq \hat{\mathcal{I}} := \int_G T^\dag (g) |\phi\rangle\langle \phi| T(g) \, 
dV(g) ,\eeq
where $dV(g)$ is a right $G$-invariant  volume element.
Because $\hat T$ is unitary, we  have the equality
\beq \hat{\mathcal{I}}\, \hat T(\alpha) = 
  \hat T(\alpha)\, \hat{\mathcal{I}} , \quad \forall\,\alpha\in G.
  \label{eq:1.resolutionId}\eeq
Therefore, by Schur's lemma,  $\hat{\mathcal{I}}$ is a multiple of the identity
 and,  with a suitable normalisation of the volume element $dV$,
 \beq     
 \int_G \Psi^*(g) \Phi(g) \, dV(g) =
\int_G \langle \phi | \hat T^\dag (g) |\phi\rangle\langle \phi|\hat T(g) |\psi\rangle\, 
dV (g) =  \langle \psi |\varphi\rangle . \label{eq:1.FundIP}\eeq

\subsection{More general coherent state representations}

The above coherent state representations are defined on Hilbert spaces of complex-valued functions for any choice of fiducial vector, $|\phi\rangle$.
However,  a judicious choice of $|\phi\rangle$ can result in major simplifications.
In particular, it is known \cite{Perelomov85,KlauderS85} that, for some choices, it is possible to identify  systems of coherent states vectors with simpler properties, that also span the Hilbert space and give rise to more useful coherent state realisations of a given irrep.  
Moreover, as we  illustrate,  evaluation of the inner products for the corresponding coherent state wave functions by  algebraic K-matrix methods  can also become much easier
and apply more generally (particularly for irreps of non-compact groups for which the above resolution of the identity does not satisfy the required convergence conditions).

Assume that the representation $\hat T$ has an extension to a representation of 
$G^\Cb$, the complex extension of $G$, defined by the natural complex extension of the Lie algebra of $G$.   
Let $|\phi\rangle$ be a fiducial vector and let $N$ be a subset of $G^c$ such that the coherent-state vectors
\beq \{ \hat T^\dag(z) |\phi\rangle ; z\in N\} \eeq
span $\Hb$.  Then, any vector $|\psi\rangle\in \Hb$ is uniquely defined by the 
coherent state wave functions
\beq \Psi(z) := \langle \phi |\hat T(z) |\psi\rangle , \quad \forall\, z\in N, \eeq
and a coherent state representation of $G$ is  defined by 
\beq \hat\Gamma(g) \Psi(z) := \langle \phi |\hat T(z) \hat T(g)|\psi\rangle , \quad \forall\, g\in G. \eeq

Such a coherent state representation is an induced representation.  
For, if $H\subset G$ denotes the  \emph{isotropy subgroup}  of all elements $h\in G$ for which
\beq \hat T(h)|\phi\rangle = \sigma(h)|\phi\rangle , 
\quad {\rm with} \; \sigma(h) \in \Cb.  \label{eq:2.isotygroup}\eeq
the map $\sigma :H\to \Cb ; h\mapsto\sigma(h)$ is the one-dimensional unitary representation of  $H$ from which the coherent state irrep 
$\hat \Gamma$ is induced.
And, as in standard induced representation theory \cite{Mackey68}, the  wave functions of the induced representation  satisfy the symmetry condition
\beq \Psi(hz) := \langle \phi |\hat T(h)\hat T(z) |\psi\rangle 
= \sigma(h)\Psi(z) , \quad \forall\, z\in N, \; h \in H. 
\label{eq:2.isotsub}\eeq

\subsection{Holomorphic coherent state representations of \Lfr{su}(1,1)}
\label{sect:su11}

Holomorphic coherent state representations can be constructed for both compact and non-compact semi-simple Lie groups and their Lie algebras.  However, the unitary irreps of non-compact groups are usually of infinite dimension, for which  extra considerations apply.  For example, whereas the irreps of compact semi-simple and reductive Lie groups have both highest and lowest weights, those of a non-compact group may have a highest or a lowest weight but, generally, not both.
A more essential distinction is that if $\{ \hat X^+_i \}$ is a set of raising operators, relative to a lowest weight state $|\phi\rangle$ for an irrep of a non-compact Lie group, the values of $\{ z_i\}$ required to define a set of coherent state vectors
\beq   \{|\phi(z)\rangle :=\exp (\sum_i z_i^* \hat X^+_i) |\phi \rangle , 
z_i \in \Cb \} ,\eeq
that span the Hilbert space for the irrep, can be restricted to a subset.
For a compact Lie group no such restriction is necessary
 because the expansion of the exponential for $|\phi(z)\rangle$ in this set
terminates when a highest weight vector  is reached.  However, when there is no highest weight state, the expansion does not terminate.
Then, for some values of  the $z_i$  variables, $|\phi(z)\rangle$  may  not converge to a normalisable vector in the Hilbert space.
Thus, the domains of the complex  variables $\{ z_i\}$ are appropriately restricted to  give subsets  of normalisable vectors, i.e., subsets for which
\beq \langle \phi(z)  |\phi(z)\rangle < \infty. \label{eq:2.zlimit}\eeq
As we now find, this complication is not much in evidence in the construction of the coherent state representation of the Lie algebra, but it has important consequences for the inner product for the Hilbert space of coherent state wave functions and their inner products (see Sect.\ \ref{sect:KmatTheory}).

We  consider  a unitary irrep $\hat T$ of \lfr{su}(1,1)  with lowest weight. 
The \lfr{su}(1,1)$^\Cb$ Lie algebra, is spanned by  a Cartan element $S_0$ and a pair of raising and lowering operators $S_\pm$ that satisfy the commutation relations
\beq [S_0, S_\pm ] = \pm S_\pm , \quad [S_-,S_+] = 2S_0 .\eeq
Let $|\lambda 0\rangle$ denote a lowest weight state for an \lfr{su}(1,1) irrep that is annihilated by  $\hat S_- := \hat T (S_-)$ and is an eigenstate of 
$\hat S_0:= \hat T(S_0)$, so that 
\beq \hat S_- |\lambda 0\rangle = 0 , \quad 2\hat S_0 |\lambda 0\rangle = \lambda |\lambda 0\rangle .
\label{eq:2.SU11lwt}\eeq
A state $|\psi\rangle$ in the Hilbert space of this irrep then has a holomorphic coherent state wave function with values
\beq \Psi^{(\lambda)}(z) =  \langle\lambda 0|e^{z \hat S_-} |\psi\rangle ,\eeq
with $z$ restricted to values for which \eq{eq:2.zlimit} is satisfied.
The coherent state representation $\hat\Gamma^{(\lambda)}$ of the \lfr{su}(1,1) algebra on these wave functions is then defined by
\beq \hat\Gamma^{(\lambda)} (X) \Psi^{(\lambda)}(z) :=
\langle\lambda 0|e^{z \hat S_-}\hat X |\psi\rangle, \quad {\rm for} \; X\in \lfr{su}(1,1)^\Cb .
\eeq
Thus, if $\hat{\mathcal{S}}^{(\lambda)}_0 := \hat\Gamma^{(\lambda)} (S_0)$ and  
$\hat{\mathcal{S}}^{(\lambda)}_\pm := \hat\Gamma^{(\lambda)} (S_\pm)$,
we obtain
\beqa \hat{\mathcal{S}}^{(\lambda)}_-\Psi (z) = \langle\lambda 0| e^{z\hat S_-} \hat S_- |\psi\rangle
=  \langle\lambda 0|\hat S_- e^{z\hat S_-}  |\psi\rangle, \\
 \hat{\mathcal{S}}^{(\lambda)}_0\Psi (z) 
 = \langle\lambda 0| e^{z\hat S_-} \hat S_0 |\psi\rangle
=  \langle\lambda 0| \big[ \hat S_0 + z \hat S_-\big] e^{z\hat S_-}  |\psi\rangle , \\
 \hat{\mathcal{S}}^{(\lambda)}_+\Psi (z) 
 = \langle\lambda 0| e^{z\hat S_-} \hat S_+ |\psi\rangle
=  \langle\lambda 0| \big[ \hat S_+ + 2z \hat S_0  + z^2 \hat S_-\big]e^{z\hat S_-}  |\psi\rangle , 
\eeqa
and the coherent state representation
\beq \hat{\mathcal{S}}^{(\lambda)}_- = \frac{\partial}{\partial z} ,\quad 
\hat{\mathcal{S}}^{(\lambda)}_0 
= \frac12\lambda +z\frac{\partial}{\partial z}  ,\quad
\hat{\mathcal{S}}^{(\lambda)}_+ =z\Big(\lambda + z \frac{\partial}{\partial z}\Big) . 
\label{eq:9.SUCSops}
\eeq

It is  seen that the coherent state wave function for the lowest weight state 
is the constant function $\Psi_{\lambda 0}(z)=1$ and that the raising operator increases the degree of a wave function by one.
Thus, in an elementary application of K-matrix methods, an orthonormal basis 
of  coherent-state wave functions, 
$\{\Psi_{\lambda n}\}$, is given for the irrep $T^{(\lambda)}$  by 
\beq \Psi_{\lambda n}(z) := K_{\lambda n}\, z^n, \quad n=0,1,2, \dots \eeq
with norm factors that remain to be determined.
It follows that
\beqa &\displaystyle \hat{\mathcal{S}}^{(\lambda)}_+ \Psi_{\lambda n}(z) 
= K_{\lambda n}(\lambda +n) z^{n+1} 
= (\lambda +n) \frac{K_{\lambda n}}{K_{\lambda, n+1}}\, \Psi_{\lambda, n+1}(z) , &\\
&\displaystyle \hat{\mathcal{S}}^{\lambda)}_- \Psi_{\lambda, n+1}(z) 
=  (n+1)\frac{K_{\lambda, n+1}}{K_{\lambda n}} \Psi_{\lambda n}(z) ,&\\
&\displaystyle \hat{\mathcal{S}}^{(\lambda)}_0 \Psi_{\lambda n}(z) 
= \textstyle \big(\frac12\lambda +n\big)\Psi_{\lambda n}(z) . &
\eeqa
The norm factors are then determined in K-matrix theory by requiring the matrices of the \lfr{su}(1,1) representation  to satisfy the Hermiticity relationships, 
\beqa \langle{\lambda,n+1}|\hat{{S}}_+ |{\lambda n}\rangle
= \langle {\lambda n}|\hat{{S}}_- |{\lambda, n+1})^*  ,\label{eq:2.HermEq1}\\ 
 \langle{\lambda n}| \hat{{S}}_0 |{\lambda n}\rangle
=\langle{\lambda n}| \hat{{S}}_0 |{\lambda n}\rangle^*, 
\label{eq:2.HermEq2} \eeqa
 required of a unitary irrep.
  To satisfy \eq{eq:2.HermEq1}, it is required that
\beq \left| \frac{K_{\lambda, n+1}}{K_{\lambda n}} \right|^2 = \frac{\lambda +n}{n+1} .
\label{eq:9.KratioSU11}\eeq
Equation (\ref{eq:2.HermEq2}) is then also satisfied and we obtain the standard expressions 
\beqa
 \hat{\mathcal{S}}_0 \Psi_{\lambda n}(z) 
 = \textstyle \big(\frac12\lambda +n\big)\Psi_{\lambda n}(z) ,    \\
\hat{\mathcal{S}}_+ \Psi_{\lambda n} 
= \sqrt{(\lambda+n)(n+1)} \, \Psi_{\lambda,n+1} ,  \\
\hat{\mathcal{S}}_- \Psi_{\lambda,n+1} 
= \sqrt{(\lambda +n)(n+1))} \, \Psi_{\lambda n} .
\eeqa

If needed, the recursion relation (\ref{eq:9.KratioSU11}) for $K_{\lambda n}$ is easily solved with $K_{\lambda 0}=1$ to give
\beq K_{\lambda n} = \sqrt{\frac{(\lambda + n-1)!}{(\lambda-1)! \, n!}} \eeq
and the orthonormal basis of coherent-state wave functions
\beq\Psi_{\lambda n}(z) = \sqrt{\frac{(\lambda + n -1)!}{(\lambda-1)!\, n!}}\,  z^n , \quad  n=0,1,2, \dots \, .   \label{eq:2.su11wfn}\eeq

\subsection{An SO(3) coherent state representations of \Lfr{su}(3)}

The scalar holomorphic coherent state representations of reductive and semi-simple Lie groups and algebras, considered above, have proved to be useful and insightful in numerous applications.  However, there are other possibilities and, for practical purposes, some that are more useful.  For SU(3), for example, the scalar holomorphic representations are limited to a subset of irreps in a canonical SU(2) basis, whereas in physical applications, especially in nuclear physics, one needs the full set of irreps in an SO(3)-coupled basis. Such coherent state irreps are readily constructed \cite{RoweLeBR89,RoweVC89}
 from the observation that, provided a highest weight state $|\lambda\mu\rangle$ for the desired SU(3) irrep 
$\hat T^{(\lambda\mu)}$ is not an eigenstate of any component of the \lfr{so}(3) angular momentum algebra, a complete set of SO(3) coherent states, i.e., a set that spans the SU(3) irrep of highest weight 
$(\lambda \mu)$, is given by the set
\beq \{ \hat T^{(\lambda\mu)}(\Omega) |\lambda\mu\rangle , 
\; \Omega \in {\rm SO}(3)\} .\eeq
We show in the following sections that complete sets of SU(3) irreps, in both canonical SU(2) and SO(3) bases, are given more usefully in VCS theory.


\section{Vector coherent state (VCS) representation} \label{sect:3.VCSdefn}

A VCS irrep is  an irrep of a Lie group $G$ that is induced from a multi-dimensional irrep of a subgroup by generalising the coherent state construction to vector-valued wave functions.
Vector-valued coherent state methods  
\cite{Rowe84a,Rowe84,RoweRC84} and  related 
{partial coherent-state methods} \cite{DeenenQ84}  were introduced
for the non-compact Sp$(n,\Rb)$ symplectic groups  in 1984. 
In fact,  holo\-morphic vector-valued representations of the Sp$(n,\Rb)$ groups had  been constructed  many years previously by Harish-Chandra 
\cite{HarishChandra55}.  
However, they were not used in physics because of the intractable nature of their inner products.
This obstacle was  resolved, within the framework of VCS theory, by  algebraic K-matrix methods \cite{Rowe84,Rowe95}  which by-pass the need for carrying out the computationally intensive  integrals of the  Harish-Chandra inner products \cite{Godement58,Gelbart73}.  They were nevertheless shown 
\cite{RoweRG85} to give the same results.
An early review of VCS theory and its applications was given by Hecht 
\cite{Hecht87}.


Basic VCS irreps are defined as follows.
Let $\hat \sigma$ denote a unitary irrep of a subgroup $H\subset G$ and suppose that this irrep is realised in the restriction of a unitary irrep $\hat T$ of $G$ to $H$.  
This means that the Hilbert space, $\Hb$, for the irrep $\hat T$ of $G$, 
contains a subspace $\Hb_0\subset \Hb$ that is $H$-invariant, i.e.,
\beq \hat T(h) |\phi\rangle \in \Hb_0 , \quad \forall\, h\in H\; {\rm and}\; \forall
|\phi\rangle \in \Hb_0 ,\eeq
and that this subspace is irreducible and equivalent to $\hat \sigma$ under the restriction of $\hat T$ to $H$.
Moreover, it is possible to choose an orthonormal basis, 
$\{ |\nu\rangle\}$ for $\Hb_0$ in such a way that the operator
\beq \hat\Pi := \sum_\nu \xi_\nu \langle \nu | \eeq
intertwines the representation $\hat \sigma$ and the irrep $\hat T(H)$ on 
$\Hb_0$, i.e., 
\beq \hat\Pi \hat T(h) = \hat\sigma (h)\hat\Pi , \quad \forall\, h\in H ,\eeq
A vector $|\psi\rangle \in \Hb$ can now be represented by a vector-valued wave function 
\beq \Psi(g) := \hat \Pi \hat T(g) |\psi\rangle 
= \sum_\nu \xi_\nu \langle \nu |\hat T(g) |\psi\rangle , \quad g\in G,
\label{eq:3.PsioverG}\eeq
which satisfies the identity
\beq \Psi(hg) = \hat \sigma(h) \Psi(g) , \quad \forall\, h\in H,\; {\rm and} \;
\forall\, g\in G. \label{eq:3.VCSconstraint}\eeq

The Hilbert space of such VCS wave functions, $\mathcal{H}$, carries a coherent-state irrep $\hat \Gamma$, induced from the irrep $\hat\sigma$ of $H$, which is isomorphic to $\hat T$ and defined by
\beq \hat\Gamma(g) \Psi(g') :=  \Psi(g'g),   \quad \forall \, g,g' \in G .  
\label{eq:3.GammaG}\eeq
Thus, $\mathcal{H}$ has an  inner product inherited from the map  
$\Hb \to \mathcal{H} ; |\psi\rangle \mapsto \Psi$. 
When   $G$ is compact, and for some (e.g., discrete series) representations when $G$ is non-compact, the inner product for $\mathcal{H}$
is defined by the resolution of the identity operator
\beq \hat{\mathcal{I}} := \int_G \sum_{\mu\nu}
T^\dag (g) |\mu\rangle \xi_\mu^\dag \cdot \xi_\nu \langle \nu| T(g) \, 
dV(g) ,\eeq
where $dV(g)$ is a right $G$-invariant  volume element.
 Because $\hat T$ is unitary, we then have the equality
\beq \hat{\mathcal{I}}\, \hat T(\alpha) = 
  \hat T(\alpha)\, \hat{\mathcal{I}} , \quad \forall\,\alpha\in G .
  \label{eq:1.resolutionIdVCS}\eeq
Therefore, by Schur's lemma,  $\hat{\mathcal{I}}$ is a multiple of the identity.
Thus,  if $\Psi$ and $\Phi$ are, respectively, VCS wave functions for vectors 
$|\psi\rangle$ and $|\phi\rangle$
in $\Hb$, defined as vector-valued functions over $G$ by \eq{eq:3.PsioverG},  it follows that, with a suitable normalisation of the volume element $dV$, 
 \beqa     
 \int_G \Psi^\dag(g) \cdot\Phi(g) \, dV(g) &=&
  \int_G \sum_{\mu\nu}
\langle \psi|T^\dag (g) |\mu\rangle \xi_\mu^\dag \cdot 
\xi_\nu \langle \nu| T(g)|\phi\rangle \, dV(g) \nonumber \\
&=& \langle \psi |\varphi\rangle . \label{eq:3.FundIP}\eeqa

The above definition of a VCS irrep is useful for formal purposes, but in practice VCS wave function are defined more usefully and more generally in terms of a subgroup or subset $N\subset G^\Cb$ such that, if the subspace $\Hb_0\subset \Hb$ is of dimension $d$, the  coherent state vectors 
\beq \{ \hat T^\dag (z) |\nu_i\rangle ; z\in N,\; i=1,\dots, d\} \eeq
span $\Hb$.  
The construction then parallels that given above except that K-matrix methods are required, as for scalar coherent state representations, to evaluate  inner products.

\section{Holomorphic  VCS irreps of the \Lfr{u}(3) Lie algebra}%
\label{sect:3.VCSU3}%

The construction of  holomorphic irreps of the \lfr{u}(3) Lie algebra by VCS methods  serves as a prototype for   parallel constructions for other semi-simple and reductive Lie algebras.
A simple application 
of the holomorphic VCS construction to a  Lie algebra \lfr{g}, requires  that the complex extension, 
$\lfr{g}^\Cb$,    of \lfr{g} can be expressed as a vector space sum of  the complex extension of a compact subalgebra $\lfr{h}\subset \lfr{g}$  plus Abelian subalgebras, $\lfr{n}_\pm$, of raising and lowering operators, i.e.,
\beq \lfr{g}^\Cb = \lfr{h}^\Cb \oplus \lfr{n}_+ \oplus \lfr{n}_- .\eeq
The only classical Lie algebras for which this is not always possible are those of the odd orthogonal groups SO$(2n+1)$, for which more general VCS constructions \cite{RoweLeBH87} are  required.
A practical limitation in the application of holomorphic VCS irreps arises because  the expressions it gives for orthonormal bases and matrix elements are  in terms of Clebsch-Gordan and Racah coefficients of the subalgebra \lfr{h}; these are currently only available  for the semi-simple Lie algebras \lfr{su}(2), \lfr{su}(3) and \lfr{so}(4), and their reductive extensions, e.g., \lfr{u}(2) and \lfr{u}(3).
Fortunately, this limitation did not exclude their application  to  the  discrete series irreps of the non-compact symplectic Lie algebra 
$\lfr{sp}(3,\Rb)$, induced from those of its maximal compact subalgebra 
\lfr{u}(3) \cite{Rowe84a,RoweRC84,Rowe84}.

The complex extension, $\lfr{u}(3)^\Cb$, of \lfr{u}(3) is the Lie algebra of  all complex $3\times 3$ matrices.  It is spanned by 9 matrices $\{C_{ij}\}$ with elements
\beq (C_{ij})_{kl} = \delta_{i,k} \delta_{j,l} \label{eq:2.Cijmatrices}\eeq
and commutation relations
\beq [  C_{ij},   C_{kl}] = \delta_{k,j}  C_{il} -\delta_{i,l}  C_{kj} .
\label{eq:2.su3CR}\eeq
The subset $\{ C_{ii}; i=1,2,3\}$ spans a
Cartan subalgebra  and the subsets $\{  C_{ik}; i<k\}$ and
$\{  C_{ik}; i>k\}$ are, respectively, raising and lowering operators.

We consider a generic \lfr{u}(3) irrep, $\hat T^{(\lambda)}$, in which the matrices 
$\{ C_{ij}\}$  are represented by operators,  $\{ \hat C_{ij}\}$, on a Hilbert space, $\Hb^{(\lambda)}$, with highest-weight  
$\lambda :=\{\lambda_1\lambda_2\lambda_3\}$, where $\lambda_1$, $\lambda_2$, and $\lambda_3$ differ by  integers and satisfy the inequality  $\lambda_1\geq\lambda_2\geq\lambda_3 $.
We then define a subspace $\Hb^{(\lambda)}_0\subset \Hb^{(\lambda)}$ of so-called \emph{highest-grade} states, defined by
\beq \Hb^{(\lambda)}_0 := \big\{ |\psi\rangle \in \ \Hb^{(\lambda)} \, \big| \,
\hat C_{12} |\psi\rangle = \hat C_{13} |\psi\rangle = 0\big\} .\eeq
The isotropy subgroup, $H$,  of all U(3) elements that leave 
$\Hb^{(\lambda)}_0$ invariant is then a group with   Lie algebra,
 $\lfr{h} =\lfr{u}(1) \oplus \lfr{u}(2)$, for which the complex extension is spanned by the zero and horizontal root vectors,  $\{ C_{11}, C_{ik},  i,k = 2,3 \}$,  of the \lfr{u}(3) root diagram, shown in Fig.\ \ref{fig:3.2}(a).
\begin{figure}[ht]
\includegraphics[scale=.23]{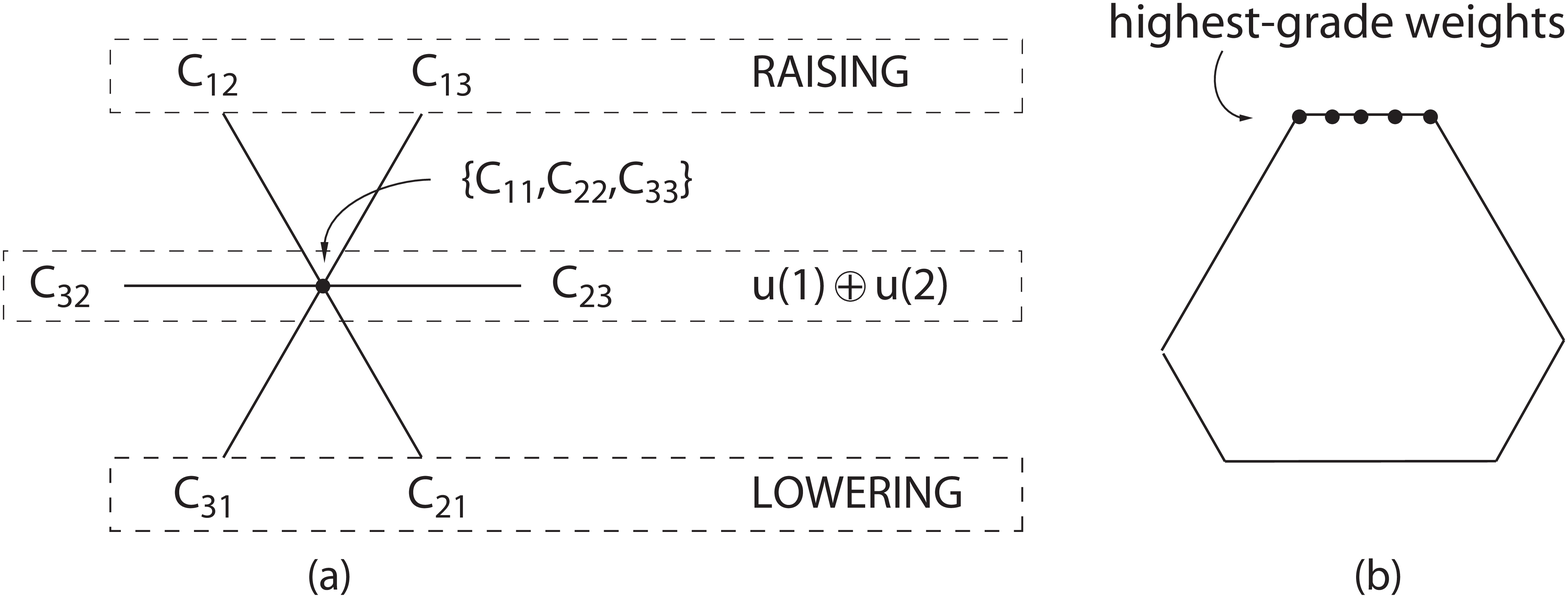}
\caption{The root diagram for \fr{u}(3), Fig.\ (a),  and the  boundary of the 	
    weight diagram for an irrep 
    $\lambda :=\{\lambda_1\lambda_2\lambda_3\}$, Fig.\ (b).  The Hermitian 
    combinations of the horizontal root vectors span the isotropy subalgebra,  
    $\fr{h} =\fr{u}(1)\!\,\oplus \!\, \fr{u}(2)\!\subset\! \fr{u}(3)$, of elements 
    that leave  invariant the space, $\Hb^{(\lambda)}_0$, spanned by the 
    highest-grade vectors whose weights are  shown.}
    \label{fig:3.2}
\end{figure}

The Hilbert space, $\Hb^{(\lambda)}_0$, carries a known irrep of the Lie algebra $\lfr{h}$.
The elements 
\beq 2 S_0 :=  C_{22} -  C_{33} , \quad
 S_+ :=  C_{23} , \quad   S_- :=  C_{32}   , \label{eq:3.qsops}\eeq
are standard basis elements for the complex extension of an \lfr{su}(2) subalgebra of $\lfr{h}$. 
Thus,  $\Hb^{(\lambda)}_0$ is spanned by  orthonormal basis vectors, 
$\{ |(\lambda)sm\rangle\}$, for which
\beqa \hat C_{11} |(\lambda)sm\rangle = \lambda_1 |(\lambda)sm\rangle , \quad
\big(\hat C_{22} + \hat C_{33}\big) |(\lambda)sm\rangle =  
\big(\lambda_2 +\lambda_3\big) |(\lambda)sm\rangle , \label{eq:g0a}\\
\hat S_0  |(\lambda)sm\rangle = m |(\lambda)sm\rangle , \\
\hat S_\pm  |(\lambda)sm\rangle = \sqrt{(s\mp m)(s\pm m +1)} 
|(\lambda)s,m\pm 1\rangle ,\\
 s= \textstyle \frac12 (\lambda_2-\lambda_3) , \quad m= -s, -s+1, \dots , s .
 \label{eq:g0b}\eeqa
Thus, the irrep  of $\lfr{h}$ carried by the highest-grade subspace  $\Hb^{(\lambda)}_0\subset\Hb^{(\lambda)}$ is completely defined by the highest-weight  $\lambda$. 

In accordance with the  principles outlined in Sect.\ \ref{sect:3.VCSdefn}, we now introduce
intrinsic wave functions, $\{\xi^{(\lambda)}_{sm}\}$, for the states
$\{ |(\lambda)sm\rangle\}$  and a corresponding irrep, $\hat \sigma$, of the
operators of $\lfr{h}$ by intrinsic operators  
$\{ \hat \sigma_{11}, \hat \sigma_{ik} , i,k = 2,3\}$.
Eqns.\ (\ref{eq:g0a}) - (\ref{eq:g0b}) then define corresponding transformations of these intrinsic wave functions:
\beqa \hat \sigma_{11} \xi^{(\lambda)}_{sm} 
= \lambda_1 \xi^{(\lambda)}_{sm} , \quad
\big(\hat \sigma_{22} + \hat \sigma_{33}\big) \xi^{(\lambda)}_{sm} =  
\big(\lambda_2 +\lambda_3\big) \xi^{(\lambda)}_{sm} , \label{eq:g0c}\\
\hat s_0  \xi^{(\lambda)}_{sm} = m \xi^{(\lambda)}_{sm} , \quad
\hat s_\pm  \xi^{(\lambda)}_{sm} 
= \sqrt{(s\mp m)(s\pm m +1)} \xi^{(\lambda)}_{s,m\pm 1} , 
 \label{eq:g0d}\eeqa
where 
\beq 2 \hat s_0 :=  \hat\sigma_{22} -  \hat\sigma_{33} , \quad
 \hat s_+ :=  \hat\sigma_{23} , \quad   \hat s_- :=  \hat\sigma_{32}   , 
 \label{eq:7.qsopsb}\eeq

A VCS irrep of \lfr{u}(3) of highest weight $\lambda$ is now induced from the irrep $\hat \sigma$ of $\lfr{h}$ in  parallel with the construction of a holomorphic scalar coherent state irrep.  In this irrep,  a vector
 $|\psi\rangle\in\Hb^{(\lambda)}$ is represented by a holomorphic vector-valued wave function 
\beq \Psi(z) := 
\sum_m \xi^{(\lambda)}_{sm} \langle {(\lambda)}sm |e^{\hat Z} |\psi\rangle , \quad
\hat Z := \sum_k z_k \hat C_{1k} .\eeq
The corresponding  representation, 
$C_{\mu\nu} \to \hat{\mathcal{C}}^{(\lambda)}_{\mu\nu}\equiv 
\hat\Gamma^{(\lambda)}(C_{\mu\nu})$, of the \lfr{u}(3) Lie algebra, is then defined in the standard way by
\beq  \hat{\mathcal{C}}^{(\lambda)}_{\mu\nu}\Psi(z) := 
\sum_m \xi^{(\lambda)}_{sm} \langle{(\lambda)} sm|  e^{\hat Z} \hat C^{(\lambda)}_{\mu\nu} |\psi\rangle. \eeq
Thus, 
with $i,k >1$ and  the expansions
\beqa \hat{\mathcal{C}}^{(\lambda)}_{1i}\Psi(z) =
  \sum_m \xi{(\lambda)}_{sm}
   \langle {(\lambda)}{sm}|\hat C_{1i}  e^{\hat Z}  |\psi\rangle, \\
 \hat{\mathcal{C}}^{(\lambda)}_{11} \Psi(z) = 
  \sum_m \xi^{(\lambda)}_{sm}  
  \langle{(\lambda)}{sm} | \big[ \hat C_{11} - \!\sum_{k=2}^3 z_k \hat C_{1k}\big] e^{\hat Z} |\psi\rangle , \\
\hat{\mathcal{C}}^{(\lambda)}_{ik} \Psi(z) =
 \sum_m \xi^{(\lambda)}_{sm}  \langle {(\lambda)}{sm} | 
 \big[ \hat C_{ik}  + z_i \hat C_{1k}\big] e^{\hat Z} |\psi\rangle , \\
 \hat{\mathcal{C}}^{(\lambda)}_{i1}\Psi(z)= 
\sum_{m} \xi^{(\lambda)}_{sm}  
\langle {(\lambda)}{sm}| \big[ z_i \hat C_{11} -\! \sum_{k=2}^3 z_k \hat C_{ik}
 - z_i \sum_{k=2}^3 z_k \hat C_{1k}\big]e^{\hat Z}  |\psi\rangle ,
\eeqa
we obtain
\beqa &&\hat{\mathcal{C}}^{(\lambda)}_{1i} = \partial_i := \partial/\partial z_i, \label{eq:3.VCSC1}\\
&&\hat{\mathcal{C}}^{(\lambda)}_{11} = \lambda_1 -\! \sum_{k=2}^3 z_k\partial_k  ,\quad
\hat{\mathcal{C}}^{(\lambda)}_{ik} = \hat\sigma_{ik} + z_i\partial_k  , 
\label{eq:3.VCSC2}\\
&&\hat{\mathcal{C}}^{(\lambda)}_{i1} = \lambda_1 z_i - 
 \sum_{k=2}^3\hat \sigma_{ik} z_k - z_i \sum_{k=2}^3 z_k\partial_k  .
 \label{eq:3.VCSC3}
 \eeqa
 Derivation of the operators of a VCS representation of any classical Lie algebra, except for the odd orthogonal algebras, is equally straightforward.  

We now consider the determination of u(3) matrix elements  in an orthonormal basis. This is simplified by use of  SU(2) tensorial methods.
With \lfr{su}(2) spin operators defined in terms of the VCS operators of 
\eq{eq:3.VCSC2} by
\beq 2\hat{\mathcal{S}}^{(\lambda)}_0 
= \hat{\mathcal{C}}^{(\lambda)}_{22}-\hat{\mathcal{C}}^{(\lambda)}_{33}, \quad
\hat{\mathcal{S}}^{(\lambda)}_+ = \hat{\mathcal{C}}^{(\lambda)}_{23}, \quad
\hat{\mathcal{S}}^{(\lambda)}_- = \hat{\mathcal{C}}^{(\lambda)}_{32}, \eeq
it is determined that the $z_2$ and $z_3$ variables obey the commutation relations
\beqa [\hat{\mathcal{S}}^{(\lambda)}_0, z_2] = \textstyle \frac12 z_2 , \quad 
[\hat{\mathcal{S}}^{(\lambda)}_0, z_3] = \textstyle -\frac12 z_3 , \\ 
{[}\hat{\mathcal{S}}^{(\lambda)}_+, z_3] =\textstyle \frac12 z_2 , \quad 
[\hat{\mathcal{S}}^{(\lambda)}_-, z_2] =\textstyle \frac12 z_3 ,. \eeqa
Thus, they transform, respectively, as the $\pm \afrac1/2$ components of an SU(2) spin-$\afrac1/2$ tensor.
The intrinsic vectors $\{ \xi^{(\lambda)}_{sm}\}$ transform as basis vectors of a spin-$s$ irrep.
Thus, it is appropriate to define an orthonormal SU(2)-coupled basis of VCS wave functions in the form
\beq \Psi^{(\lambda)}_{jSM}(z) := K^{(\lambda)}_{jS} \big[  \varphi_{j}(z)
\otimes \xi_s \big]_{SM} ,     \label{eq:3.PsisjS}     \eeq
where
\beq \varphi_{jm}(z) := \frac{z_2^{j+m}z_3^{j-m}}{\sqrt{(j+m)! (j-m)!}} \,,
\quad m = -j, -j+1, \dots, +j , \label{eq:3.phijm}\eeq
$[  \varphi_{j}(z)\otimes \xi_s ]_{SM} := \textstyle\sum_{m\nu}
(s\nu, jm |SM)\, \varphi_{jm}(z)\otimes \xi_{s\nu}$, where 
$(s\nu, jm |SM)$ is an SU(2) Clebsch-Gordan coefficient,
and $\{ K^{(\lambda)}_{jS}\}$ is a set of norm factors  to be determined.
These basis wave functions then span subsets of \lfr{su}(2) irreps for which
\beqa
\hat{\mathcal{S}}^{(\lambda)}_0\Psi^{(\lambda)}_{jSM} 
= M \Psi^{(\lambda)}_{jSM} ,  \\
\hat{\mathcal{S}}^{(\lambda)}_\pm\Psi^{(\lambda)}_{jSM} 
= \sqrt{(S\mp M)(S\pm M+1)}\,   \Psi^{(\lambda)}_{jS,M\pm 1} .
\eeqa
The actions of  elements of $\lfr{h}$ are similarly given by
\beqa
\hat{\mathcal{C}}^{(\lambda)}_{11}  \Psi^{(\lambda)}_{jSM} 
= \big(\lambda_1 -2j\big)  \Psi^{(\lambda)}_{jSM} , \label{eq:4.grade} \\
(\hat{\mathcal{C}}^{(\lambda)}_{22} + \hat{\mathcal{C}}^{(\lambda)}_{33} )
 \Psi^{(\lambda)}_{jSM}, =
\big(\lambda_2 + \lambda_3  +2j\big)  \Psi^{(\lambda)}_{jSM}. \label{eq:4.grade2}
\eeqa

It remains to determine the matrix elements of the raising and lowering operators, $\{ \hat C_{1i}\}$ and $\{ \hat C_{i1}\}$, between bases of different $j$ and $S$.  These  operators are components of SU(2) spin-$\afrac1/2$ tensors:
\beqa \hat e_{\safrac1/2} :=  \hat C_{13}, &\quad & 
\hat e_{-\safrac1/2} := - \hat C_{12}, \\
\hat f_{\safrac1/2} :=  \hat C_{21}, &\quad & 
\hat f_{-\safrac1/2} :=  \hat C_{31} ,
\eeqa
consistent with the relationship 
$\big(\hat e_m\big)^\dag = (-1)^{\safrac1/2 -m} \hat f_{-m}$.
 In the VCS representation, given by Eqns.\ 
 (\ref{eq:3.VCSC1}) -  (\ref{eq:3.VCSC3}),
the raising operators are mapped to spin-$\afrac1/2$ differential operators:
\beq \hat e_{\pm\!\safrac1/2} \to \hat d_{\pm\!\safrac1/2} ,\eeq
where $\hat d_{\safrac1/2} := \partial_3$ and $\hat d_{-\safrac1/2} := -\partial_2$.
Thus, their SU(2)-reduced matrix elements are easily derived.  The VCS expressions for the lowering operators, given by \eq{eq:3.VCSC3},
are seemingly more complicated.  
In fact, their VCS representation is expressed almost as simply in the form 
\beq \hat f_{\pm\!\safrac1/2} \to [\hat\Lambda ,\hat z_{\pm\!\safrac1/2}] , 
\label{eq:7.Fa} 
\eeq
where $\hat z_{\safrac1/2} := z_2$, $\hat z_{-\!\safrac1/2} := z_3$, and
$\hat\Lambda$ is the U(2)-scalar operator
\beq \hat\Lambda := \lambda_1 \sum_{i=2}^3 z_i\partial_i - \sum_{i,k=2}^3 \hat\sigma_{ik} z_k \partial_i - {\t\frac12} \sum_{i,k=2}^3 z_i z_k \partial_k \partial_i  .\label{eq:7.Lambda}\eeq
Expressing the lowering operators in this way 
 (which is standard in VCS theory \cite{Rowe84})
 greatly facilitates the calculation of their matrix elements because the operator
$\hat \Lambda$ is a multiple of the identity within a U(2) irrep. Consequently,  
it is  diagonal in the above-defined SU(2)-coupled VCS basis, i.e.,  
\beq  \hat\Lambda \Psi^{(\lambda)}_{jSM} 
= \Omega^{(\lambda)}_{jS} \Psi^{(\lambda)}_{jSM} ,\eeq
 with eigenvalues given for an irrep with highest weight 
 $\lambda \equiv \{\lambda_1\lambda_2\lambda_3\}$ by
\beq \Omega^{(\lambda)}_{jS} = \big(2\lambda_1-\lambda_2-\lambda_3\big) j -S(S+1) + s(s+1) -j(j-2) .  \label{eq:3.Fa}\eeq

Matrix elements of the $\hat f$ operators are now calculated as follows.
With $\Psi^{(\lambda)}_{jSM}$ defined by \eq{eq:3.PsisjS}, it follows that
\beq \big[ \hat  z\otimes \Psi^{(\lambda)}_{jS}\big]_{S'M'} 
= K^{(\lambda)}_{jS} \big[\hat  z\otimes [ \varphi_j \otimes \xi_s]_S\big]_{S'M'} \eeq
and, with some Racah recoupling, we obtain
\beq  \big[\hat   z\otimes \Psi^{(\lambda)}_{jS}\big]_{S'M'} 
=\sum_{j'}  K^{(\lambda)}_{jS}\, 
 U(sjS' \afrac1/2 : Sj') 
\big[  [ \hat  z\otimes \varphi_j ]_{j'}\otimes \xi_s \big]_{S'M'} .  
\eeq
From the explicit expression for $\varphi_{jm}$, given by \eq{eq:3.phijm}, it is then determined that
\beq [ \hat  z\otimes \varphi_j ]_{j'm'}  = \delta_{j', j+\!\safrac1/2} \sqrt{2j+1}\, \varphi_{j+\!\safrac1/2,m'} \eeq
and, hence, that
\beq  \big[ \hat  z\otimes \Psi^{(\lambda)}_{jS}\big]_{S'M'} 
= \frac{K^{(\lambda)}_{jS}}{K^{(\lambda)}_{j+\!\safrac1/2, S'}}\,   
\sqrt{2j+1}\, U(sjS' \afrac1/2 : S, j+\!\afrac1/2) 
\Psi^{(\lambda)}_{ j+\!\safrac1/2, S'M'} .\eeq
Now, from the expression of the Wigner-Eckart theorem, in the form
\beq \big[ \hat  z\otimes \Psi^{(\lambda)}_{jS}\big]_{S'M'} 
= \sum_{j'}  \Psi^{(\lambda)}_{j'S'M'}
\frac{\langle \Psi^{(\lambda)}_{j'S'} \| \hat z \|\Psi^{(\lambda)}_{jS}\rangle}
{\sqrt{2S'+1}},
\eeq
it follows that
\beqa \langle \Psi^{(\lambda)}_{j'S'} \| \hat z \|\Psi^{(\lambda)}_{jS}\rangle &=& 
\delta_{j',j+\safrac1/2} 
\frac{K^{(\lambda)}_{jS}}{K^{(\lambda)}_{j+\safrac1/2, S'}} \, \sqrt{(2j+1)(2S'+1)} \nonumber \\
&& \times  U(sjS' \afrac1/2 : S, j\!+\!\afrac1/2) 
\eeqa
and that
\beqa 
\langle {(\lambda) j'S'} \| \hat f \|(\lambda){jS}) &=& 
\langle \Psi^{(\lambda)}_{j'S'} \|[\hat\Lambda, \hat z] 
\|\Psi^{(\lambda)}_{jS}\rangle \nonumber\\
&=& \big( \Omega_{j'S'} - \Omega_{jS} \big) 
\langle \Psi^{(\lambda)}_{j'S'} \| \hat z \|\Psi^{(\lambda)}_{jS}\rangle .
\eeqa
Thus, we obtain the reduced matrix elements
\beqa 
&&\langle {(\lambda)j'S'} \| \hat f \|(\lambda){jS}\rangle   
=  \delta_{j', j+\!\safrac1/2}
\frac{K^{(\lambda)}_{jS}}{K^{(\lambda)}_{j+\!\safrac1/2, S'}} 
\big( \Omega_{j+\!\safrac1/2,S'} - \Omega_{jS} \big)\, \nonumber \\
&&\qquad\qquad \qquad \times  \sqrt{(2j+1)(2S'+1)}\  U(sjS' \afrac1/2 : S, j\!+\!\afrac1/2)
 . \label{eq:7.Fb} \eeqa

Starting from the equation
\beq \big( \hat d \otimes \varphi_j \big)_{j'}  = -\delta_{j', j-\!\safrac1/2} \sqrt{2j+1}\, \varphi_{j-\!\safrac1/2} ,\eeq
reduced matrix elements of the raising operator tensor, $\hat e$, are similarly determined to be given by
\beqa 
&&\langle{(\lambda)jS} \| \hat e \|(\lambda){j'S'}\rangle   
=  (-1)^{S'-S %
+ %
\safrac1/2} 
\frac{K^{(\lambda)}_{j+\!\safrac1/2, S'}}{K^{(\lambda)}_{jS}}
\delta_{j', j+\!\safrac1/2} \, \nonumber \\
&&\qquad\qquad\qquad  \times  \sqrt{(2j+1)(2S'+1)}\  U(sjS' \afrac1/2 : S, j\!+\!\afrac1/2)
 . \label{eq:7.Fc} \eeqa
Finally, the norm ratios are determined from the Hermiticity condition
\beqa 
&&{\langle(\lambda)  j\!+\!\afrac1/2,\!S'\!, \!M\!+\!\afrac1/2 | \hat f _{\!\safrac1/2}  
|(\lambda)jSM\rangle^* }  \nonumber\\
&& \qquad \quad=
- \langle (\lambda)jSM |\hat e_{-\!\safrac1/2} 
 | (\lambda){j\!+\!\afrac1/2,\!S'\!, \!M\!+\!\afrac1/2}) \eeqa
which must be satisfied for the irrep to be unitary.
In terms of reduced matrix elements, this condition becomes
\beq  
\langle (\lambda){j\!+\!\afrac1/2,\!S'} \| \hat f \| (\lambda){jS}\rangle^*  
=   (-1)^{S'-S+\!\afrac1/2}
\langle(\lambda){jS} \| \hat e \|(\lambda) {j\!+\!\afrac1/2,\!S'}) .\eeq
Together with  Eqns.\ (\ref{eq:7.Fa}),  (\ref{eq:7.Fb}), and (\ref{eq:7.Fc}), this condition implies that the norm factors must satisfy the identity
\beqa 
\left| \frac{K^{(\lambda)}_{j+\!\safrac1/2, S'}} {K^{(\lambda)}_{jS}}\right|^2 
&=&  \big( \Omega^{(\lambda)}_{j+\!\safrac1/2,S'} 
- \Omega^{(\lambda)}_{jS}\big) \nonumber\\
&=& {\textstyle\frac12} (2\lambda_1-\lambda_2-\lambda_3) +S(S+1)-S'(S'+1) - j + {\textstyle \frac{3}{4} }.
\eeqa
Thus, we obtain the explicit expression for the matrix elements of a generic su(3) irrep:
\beqa 
&& \langle(\lambda){j\! +\! \afrac1/2,\! S'} \| \hat f \| (\lambda){jS}\rangle
=  (-1)^{S'-S+\afrac1/2}\, 
\langle (\lambda){jS} \| \hat e \| (\lambda){j+\!\afrac1/2,\! S'}\rangle 
\ \nonumber\\
&& \qquad\quad = \sqrt{ (2j+1)(2S'+1)} \,  U(sjS' \afrac1/2 : S,\!  j\!+\!\afrac1/2)  
\nonumber \label{eq:7.VCSu(3)}\\
&& \qquad\quad \phantom{=}  \times 
\left[{\textstyle\frac12} (2\lambda_1-\lambda_2-\lambda_3) +S(S+1)-S'(S'+1) - j + 
{\textstyle \frac34}\right]^{\frac12} , 
\eeqa
with $s= \frac12 (\lambda_2-\lambda_3)$.

Note that the above VCS construction gives analytical expressions for the matrix elements of the u(3) Lie algebra in a basis for any U(3) irrep  that reduces the subgroup chain with the associated representation labels
\beq \begin{array}{cccccccccc}
&{\rm U(3)}&\supset\quad &{\rm SU(3)}& \quad\supset\quad &{\rm U(2)} &\quad\supset\quad& {\rm U(1)}&\\
&\{\lambda_1\lambda_2\lambda_3\}& & (\lambda\mu)& &j\;S&&M&
\end{array}
,\eeq
where the SU(3) irrep labels are given by 
$\lambda = \lambda_1-\lambda_2$, $\mu = \lambda_2-\lambda_3 = 2s$.
The irreps of the subgroup ${\rm U(2)}\subset {\rm SU(3)}$ in the chain are labelled by the eigenvalues
of the U(1) operator $2\hat C_{11}-\hat C_{22}- \hat C_{33}$ which in the VCS representation are given by $(2\lambda_1-\lambda_2-\lambda_3 -6j)$, and by the SU(2) spin quantum number $S$.  A notable property of this construction is that the extra label $j$ is algebraically defined and, because of \eq{eq:4.grade2}, it defines a canonical basis for U(3) labelled by 
  well-defined quantum numbers that take integer, or half-odd integer, values.   
Such a basis is, in fact, a Gelfand-Tsetlin basis \cite{GelfandTsetlin50}.

\section{VCS irreps of \Lfr{su}(3) in an SO(3) basis} \label{sect:su3SO3}

A different class of coherent state representation  to those of the holomorphic kind was introduced in 1989 \cite{RoweLeBR89,RoweVC89}.  Its purpose was to construct the irreps of \lfr{su}(3) in an SO(3)-coupled basis, which is the basis appropriate for application to rotationally-invariant systems such  as nuclei.
A secondary purpose was to elucidate the relationship of Elliott's SU(3) model \cite{Elliott58,Elliott58b} to the non-compact nuclear rotor model that is obtained from the SU(3) model in a contraction limit.

The algebraic structure underlying  the rotor model  \cite{Ui70a} is that of a semi-direct product group with an Abelian normal subgroup.  The irreps of such a group can be induced, in an SO(3) basis,  from  one-dimensional irreps of the normal subgroup by  scalar coherent state methods  that  parallel those of Mackey's theory of induced representations \cite{Mackey68}.  Thus, it was natural to induce  scalar coherent state representations of \lfr{su}(3) in an SO(3) basis.
Such a construction was subsequently applied to induce a subset of irreps of SO(5) in an SO(3) basis \cite{RoweH95} and later extended to generic VCS irreps of SO(5)   \cite{TurnerRR06}.

In this section we give the VCS irreps of \lfr{su}(3) in an SO(3) basis which are, in many respects, simpler than their scalar counterparts and raise the possibility of other similar constructions.  In particular, we give a construction in a \emph{canonical} SO(3) basis; i.e., a basis labelled by a complete set of   well-defined   quantum numbers.

In an SO(3) angular-momentum coupled basis, the \lfr{su}(3)$^\Cb$  
 algebra is spanned by three components of angular momentum
\beq  L_{0} = -{\rm i} ( C_{23}- C_{32}), \quad
 L_\pm =  {\rm i}( C_{13}- C_{ 31}) \pm ( C_{12}-
C_{21}), \label{eq:6.Ldef}\eeq
and five quadrupole moments
\beqa &  {\cal Q}^{(2)}_0 = 2h_1+h_2, \\
& {\cal Q}^{(2)}_{\pm 1} = \mp \sqrt{\textstyle\frac32}
\,[ C_{12}+ C_{21}\pm {\rm i} ( C_{13}+ C_{31})]
,\label{eq:6.Qdef}\\
& {\cal Q}^{(2)}_{\pm 2} =\sqrt{\textstyle\frac32}
\,[ h_2\pm {\rm i} ( C_{23}+ C_{32})]
,\eeqa
where $h_1 = C_{11} - C_{22}$ and $h_2 = C_{22}-C_{33}$.

\subsection{Irreps of an intrinsic \Lfr{u}(2) subalgebra}

The above expressions show that the  \lfr{su}(3) elements $L_0$ and 
${\cal Q}^{(2)}_{\pm 2}$ involve only $\{ C_{23}, C_{32}, C_{22}-C_{33}\}$.  It follows that they span an 
$\lfr{su}(2)^\Cb \subset \lfr{su}(3)^\Cb$    
subalgebra with commutation relations
\beq [L_0, {\cal Q}^{(2)}_{\pm 2}] = \pm 2  {\cal Q}^{(2)}_{\pm 2} ,\quad
[{\cal Q}^{(2)}_2,{\cal Q}^{(2)}_{-2}] = 6 L_0 . \label{eq:su2CRs}
\eeq
In addition,  ${\cal Q}^{(2)}_0 = 2h_1+h_2$ commutes with all elements of this 
\lfr{su}(2)$^\Cb$   
 subalgebra and with them spans a \lfr{u}(2)$^\Cb$ 
  subalgebra of \lfr{su}(3)$^\Cb$
  and a corresponding   \lfr{su}(2) $\subset$ \lfr{su}(3) subalgebra.

The following will show that a desired \lfr{su}(3) irrep $\hat T^{(\lambda\mu)}$, of  highest weight $(\lambda\mu)$, can be induced from an irrep $\hat \sigma$ of this $\lfr{u}(2)$ 
subalgebra, provided $\hat\sigma$ is the irrep carried by a space of highest grade states, e.g.,  states, in the Hilbert space of the irrep 
$\hat T^{(\lambda\mu)}$, that are annihilated by the
$\hat C_{12}$ and $\hat C_{13}$ raising operators, where 
$\hat C_{ij} := \hat T^{(\lambda\mu)}(C_{ij})$.
Thus, we seek a basis of highest grade states that satisfy the equations
\beqa \hat C_{12}|(\lambda\mu)K\rangle 
= \hat C_{13}|(\lambda\mu)K\rangle=0 ,   
\quad \hat L_0 |(\lambda\mu)K\rangle = K|(\lambda\mu)K\rangle, \\
\hat {\cal Q}^{(2)}_0|(\lambda\mu)K\rangle = 
(2\hat h_1+\hat h_2)|(\lambda\mu)K\rangle = 
(2\lambda+\mu) |(\lambda\mu)K\rangle .
\label{eq:6.Q0} 
\eeqa

Comparison of the commutation relations of \eq{eq:su2CRs} with the standard 
su(2)  relations 
\beq [\hat\sigma_0, \hat\sigma_\pm ] = \pm \hat\sigma_\pm ,
\quad [\hat\sigma_+,\hat\sigma_-] = 2\hat\sigma_0 .
\eeq
and  a knowledge of the \lfr{su}(2) representations, implies that
\beq  \hat {\cal Q}^{(2)}_{\pm 2}|(\lambda\mu)K\rangle
= \textstyle\sqrt{{3\over 2} (\mu \mp K)(\mu\pm K+2)}\,|(\lambda\mu)K\pm 2\rangle
\eeq
with $K$ running over the range $-\mu, -\mu+2, \dots , +\mu$.
Thus, it is convenient to define wave functions $\{\xi^{(\lambda\mu)}_K\}$ for
the highest grade states 
$\{ |(\lambda\mu)K\rangle\}$ and a corresponding  representation of the
 su(2) algebra 
\beq L_0 \to \hat \sigma(L_0) := 2\hat s_0 , \quad 
{\cal Q}^{(2)}_{\pm 2} \to  \hat \sigma(Q^{(2)}_{\pm 2}) :=    \sqrt{6} \, 
\hat s_\pm \label{eq:6.Qsigma}
\eeq
such that
\beq  \hat s_0 \xi^{(\lambda\mu)}_K = \t\frac12 K \,\xi^{(\lambda\mu)}_K , \quad
\hat s_\pm \xi^{(\lambda\mu)}_K 
= \frac12 \sqrt{(\mu \mp K)(\mu\pm K+2)}\,\xi^{(\lambda\mu)}_{K\pm 2} .
\eeq
We  refer to the wave functions $\{\xi^{(\lambda\mu)}_K\}$ as 
\emph{intrinsic spin} wave functions and to $\{\hat s_k\}$ as
intrinsic spin operators.

\subsection{The VCS   irrep of \lfr{su}(3)}
\label{sect:6.SO3VCSrep}

The  VCS irrep under construction depends critically on the following observation.
Because the angular momentum operators $\hat L_\pm$ are linear 
combinations   
 of  $\hat C_{13}-\hat C_{31}$ and  
$\hat C_{12}-\hat C_{21}$, their repeated application to the highest grade vectors of an \lfr{su}(3) irrep, as defined above, generates a complete basis for the irrep.
Thus, if $\hat R(\Omega) =\hat T^{(\lambda\mu)}(\Omega)$ denotes the representation of an SO(3) element $\Omega\in {\rm SU(3)}$, 
the set of SO(3) coherent states 
\beq \{\hat R(\Omega) |(\lambda\mu)K\rangle, \Omega\in {\rm SO(3)}, K=
-\mu, -\mu+2, \dots , +\mu \} \eeq
spans the Hilbert space for the SU(3) irrep $\hat T^{(\lambda\mu)}$.

It follows that, if $\Hb^{(\lambda\mu)}$ is the Hilbert space for the irrep 
$\hat T^{(\lambda\mu)}$, any vector $|\psi\rangle \in \Hb^{(\lambda\mu)}$ is defined by the  overlaps
$\{\langle (\lambda\mu)K |\hat R(\Omega) |\psi\rangle ,\Omega \in{\rm SO(3)}\}$.  
Also a VCS wave function $\Psi$ for the vector $|\psi\rangle$ is defined by 
\beq \Psi(\Omega) :=\sum_K\xi^{(\lambda\mu)}_K \langle (\lambda\mu)K| \hat
R(\Omega) |\psi\rangle , \quad \Omega\in {\rm SO}(3).
\eeq
The representation of an element $X$ of the su(3) algebra as an operator 
on these wave functions is then defined by
\beq [\hat \Gamma(X) \Psi] (\Omega)
:=\sum_K\xi^{(\lambda\mu)}_K 
\langle (\lambda\mu)K | \hat R(\Omega) \hat X|\psi\rangle ,
\quad \Omega \in {\rm SO}(3).
\eeq

The transformation of an angular-momentum coupled vector under a rotation is expressed in the standard way by
\beq \hat R(\Omega) |(\lambda\mu)\alpha LM\rangle =
\sum_N |(\lambda\mu)\alpha LN\rangle \scrD^L_{NM}(\Omega) ,
\quad \Omega \in {\rm SO}(3).
\eeq
where $\scrD^L_{NM}(\Omega)$ is a \ Wigner rotation matrix.
Thus, the VCS wave function of the vector $|(\lambda\mu)\alpha LM\rangle$ is
given by
\beq  \Psi^{(\lambda\mu)}_{\alpha LM}(\Omega) =\sum_K\xi^{(\lambda\mu)}_K 
\langle(\lambda\mu)K|(\lambda\mu)\alpha LK\rangle\, \scrD^L_{KM}(\Omega) ,
\quad \Omega \in {\rm SO}(3). \label{eq:5.basis1}
\eeq

The angular momentum operators of the VCS representation act on such
wave functions in the standard way
\beqa  [\Gamma(L_0) \Psi^{(\lambda\mu)}_{\alpha LM}] (\Omega) =
M\Psi^{(\lambda\mu)}_{\alpha LM}(\Omega) , \label{eq:5.L0}\\
{[}\Gamma(L_\pm) \Psi^{(\lambda\mu)}_{\alpha LM}] (\Omega) =
\sqrt{(L\mp M)(L\pm M+1)} \, \Psi^{(\lambda\mu)}_{\alpha L,M\pm 1}  (\Omega) .
\label{eq:5.Lpm}\eeqa
For the quadrupole operators, 
\beqa
&& [\Gamma({\cal Q}^{(2)}_\nu)\Psi^{(\lambda\mu)}_{\alpha LM}] (\Omega)
=\sum_K\xi^{(\lambda\mu)}_K \langle (\lambda\mu)K|\hat R(\Omega)  
\hat {\cal Q}^{(2)}_{\nu} |(\lambda\mu)\alpha LM\rangle ,\nonumber \\
&& \qquad\qquad\quad = \sum_{K\nu'}  \xi^{(\lambda\mu)}_K 
\langle (\lambda\mu)K|\hat {\cal Q}^{(2)}_{\nu'} 
\hat R(\Omega)   |(\lambda\mu)\alpha LK\rangle\, \scrD^2_{\nu'\nu}(\Omega). \label{eq:6.VCSQ}
\eeqa
Equations (\ref{eq:6.Ldef})-(\ref{eq:6.Q0}) and (\ref{eq:6.Qsigma}) then let us make the substitutions
\beqa \langle (\lambda\mu)K |\hat {\cal Q}^{(2)}_{0} =\langle (\lambda\mu)K
| (2\hat h_i + \hat h_2) = (2\lambda + \mu) \langle (\lambda\mu)K | ,\\
\langle (\lambda\mu)K |\hat {\cal Q}^{(2)}_{\pm 1} = \textstyle
-\sqrt{3\over 2}\langle (\lambda\mu)K |\hat L_\pm ,\\
\sum_K\xi^{(\lambda\mu)}_K \langle (\lambda\mu)K |\hat {\cal Q}^{(2)}_{\pm 2}
=\sqrt{6}\, \hat\sigma_\pm \sum_K\xi^{(\lambda\mu)}_K \langle (\lambda\mu)K | ,
\eeqa 
and obtain
\beqa \sum_K\xi^{(\lambda\mu)}_K \langle (\lambda\mu)K |\hat {\cal Q}^{(2)}_{0}
\hat R(\Omega)  |(\lambda\mu)\alpha LM\rangle =(2\lambda+\mu) \Psi_{\alpha
LM}(\Omega) 
 ,  \label{eq:6.95}\\
\sum_K\xi^{(\lambda\mu)}_K \langle (\lambda\mu)K |\hat {\cal Q}^{(2)}_{\pm 1}
\hat R(\Omega)  |(\lambda\mu)\alpha LM\rangle =\textstyle -\sqrt{3\over 2}\,  \big[\bar L_\pm \Psi_{\alpha LM}\big](\Omega) ,
\\ 
\sum_K\xi^{(\lambda\mu)}_K \langle (\lambda\mu)K |\hat {\cal Q}^{(2)}_{\pm 2}
\hat R(\Omega)  |\alpha LM\rangle
=\sqrt{6}\, \hat\sigma_\pm \Psi_{\alpha LM}(\Omega) ,\label{eq:6.97}
\eeqa
where  $\bar L_\pm$ are  infinitesimal generators of left
rotations.
Their actions, defined by 
\beq \big[\bar L_k \scrD^L_{KM}\big](\Omega) = \langle LK|\hat L_k \hat
R(\Omega) |LM\rangle = \sum_N \langle LK |\hat L_k |LN\rangle {\cal
D}^L_{NM}(\Omega )
\eeq
give the expressions, familiar in the nuclear rotor model,
\beqa \bar L_0 \scrD^L_{KM} = K \scrD^L_{KM} ,\\
 \bar L_\pm \scrD^L_{KM} = \sqrt{(L\pm K)(L\mp K+1)}\, \scrD^L_{K\mp 1,M}. 
\eeqa
We conclude from Eqns.~(\ref{eq:6.VCSQ}) and (\ref{eq:6.95})-(\ref{eq:6.97})
that $\hat\Gamma({\cal Q}_\nu^{(2)})$ can be expressed
\beqa \hat\Gamma({\cal Q}_\nu^{(2)}) &=& 
(2\lambda+\mu) \hat\scrD^2_{0\nu} -
\textstyle \sqrt{3\over 2}\, \big( \hat\scrD^2_{1\nu}\bar L_+ +
\hat\scrD^2_{-1\nu} \bar L_-\big) \nonumber\\
&& +\sqrt{6}\big[ \hat\sigma_+\hat\scrD^2_{2\nu} 
+ \hat\sigma_-\hat\scrD^2_{-2\nu}\big] , \label{eq:6.express1}
\eeqa
with the understanding that, as an operator $\hat\scrD^2_{\mu\nu}$ acts
multiplicatively;
\beq [\hat\scrD^2_{\mu\nu} \Psi](\Omega)= \scrD^2_{\mu\nu}(\Omega)\,
\Psi(\Omega) .
\eeq

Equation (\ref{eq:6.express1}) can be simplified by means of the identity
\beq \big[ \hat{\bf L}^2, \hat\scrD^2_{0\nu}\big] = 6\hat\scrD^2_{0\nu}+
\textstyle 2 \sqrt{3\over 2}\, 
\big[ \hat\scrD^2_{1\nu}\bar L_+ +\hat\scrD^2_{-1\nu}\bar L_-\big]
\eeq
to the more useful expression
\beq \hat\Gamma({\cal Q}_\nu^{(2)}) = (2\lambda+\mu+3) \hat\scrD^2_{0\nu} -
\textstyle
\half\,\big[ \hat{\bf L}^2, \hat\scrD^2_{0\nu}\big]
+ \sqrt{6}\big[ \hat\sigma_+\hat\scrD^2_{2\nu} +
 \hat\sigma_-\hat\scrD^2_{-2\nu}\big] . \label{eq:6.GammaQ}
\eeq

\subsection{Basis wave functions}\label{sect:6.VCSbasis}

Equation (\ref{eq:5.basis1}) shows that a basis of VCS wave functions for the 
\lfr{su}(3) irrep is given by linear combinations of the vector-valued functions
$\xi^{(\lambda\mu)}_K \scrD^L_{KM}$,
with $K= -\mu, -\mu+2, \dots , \mu$.  
The allowed combinations are further restricted by the constraint of 
\eq{eq:3.VCSconstraint} which requires that VCS wave functions should satisfy the equality
\beq  \Psi^{(\lambda\mu)}_{\alpha LM}(\Omega)
 =\sum_K \hat\sigma(\omega)\xi^{(\lambda\mu)}_K 
\langle(\lambda\mu)K|\hat R(\omega^{-1})\hat R(\Omega)
|(\lambda\mu)\alpha LK\rangle 
,\label{eq:5.constraint}
\eeq
for all $\omega\in {\rm SO(3)}$ that leave the highest-grade subspace, spanned by the  vectors $\{ |(\lambda\mu) K\rangle\}$, invariant.  
The \emph{isotropy} subgroup of such $\omega\in {\rm SO(3)}$ clearly includes the SO(2) subgroup with infinitesimal generator $L_0$, for which
\beq \hat\sigma(e^{-{\rm i}\phi L_0}) \xi^{(\lambda\mu)}_K  = 
e^{-{\rm i} K \phi } \xi^{(\lambda\mu)}_K ,
\quad \hat R(e^{{\rm i}\phi L_0}) |(\lambda\mu)K\rangle 
= e^{{\rm i}K\phi}|(\lambda\mu)K\rangle ,
\eeq
and for which the constraint condition is automatically satisfied.
However, it also contains the rotation  through angle $\pi$ generated by the angular momentum operator 
$L_y= - \frac12 {\rm i} (L_+ - L_-)$, for which 
\beqa \exp (-{\rm i}\pi L_y ) L_0 \exp ({\rm i}\pi L_y ) = -L_0 , \\
\exp (-{\rm i}\pi L_y ) Q^{(2)}_{\pm 2} \exp ({\rm i}\pi L_y ) = Q_{\pm 2} .
\eeqa

By explicit construction of the vectors $\{ \xi^{(\lambda\mu)}_K\}$ in the space of a two-dimensional harmonic oscilator, it is determined that, with 
$\omega = \exp (-{\rm i}\pi L_y )$,
\beq \hat \sigma(\omega)\xi^{(\lambda\mu)}_K =
(-1)^{\lambda}\xi^{(\lambda\mu)}_{-K}.
\eeq
It is also known that
\beq \scrD^L_{KM}(\omega^{-1}\Omega) = (-1)^{L+K} \scrD^L_{-K,M}(\Omega).
\eeq
Thus, it is determined that a basis of VCS wave functions for an \lfr{su}(3) irrep is given by linear combinations of the vector-valued functions
\beq \varphi^{(\lambda\mu)}_{KLM} =
{1\over \sqrt{2(1+\delta_{K0})}} \,\big( \xi^{(\lambda\mu)}_K\scrD^L_{KM} +
(-1)^{\lambda+L+K} \xi^{(\lambda\mu)}_{-K} \scrD^L_{-K,M}\big),
\label{eq:6.RMwfns}
\eeq
with $K\geq 0$ in the range $\mu, \mu -2, \dots,$ 1 or 0.
Such wave functions are  familiar in nuclear physics in the context of the rotor model.

\subsection{Matrix elements in an orthonormal basis}

By construction, the representation of the SO(3) subgroup of the above-defined VCS representation, with matrix elements of the \lfr{so}(3) angular momentum operators given by Eqns. (\ref{eq:5.L0}) and (\ref{eq:5.Lpm}), is already unitary.
However, the matrices of the  $\hat \Gamma ({\cal Q}^{(2)}_\nu)$ operators do not, in general, satisfy the Hermiticity relationships required of a unitary representation.  
Thus, we focus on the matrices of these operators.

With a coupled product of SO(3)  tensors defined by
\beq [A_{L_2} \otimes B_{L_1}]_{LM} := 
\sum_{M_1M_2} (L_1M_1, L_2M_2 | L M)\, 
A_{L_2M_2} \otimes B_{L_1M_1} , \eeq
where $(L_1M_1, L_2M_2 | L M)$ is an SO(3) Clebsch-Gordan coupling coefficient, and the well-known expression \cite{Rose57} for rotation matrices
\beq 
\left[ \scrD^{L_2}_{K_2}(\Omega) \otimes \scrD^{L_1}_{K_1}(\Omega)\right]_{LM}
= \sum_K (L_1K_1, L_2K_2 |LK) \scrD^L_{KM}(\Omega) ,\eeq 
it is determined from
\eq{eq:6.GammaQ} and the definition (\ref{eq:6.RMwfns}) that
\beq [\Gamma({\cal Q}^{(2)}) \otimes \varphi^{(\lambda\mu)}_{KL}]_{L'M'}
= \sum_{K'}\varphi^{(\lambda\mu)}_{K'L'}\, M^{L'L}_{K'K },
\label{eq:6.Qaction}\eeq 
with
\beqa M^{L'L}_{KK } &\!\!=\!\!& \left[ (2\lambda\! +\!\mu+3) 
-\t\half L'(L'+1) +\half L(L+1)\right] (LK ,20|L'K )
\nonumber\\ 
&&+ \delta_{K,1}\,  (-1)^{\lambda +L+1}
\textstyle\sqrt{{3\over 2}} (\mu+1)(L,-1,22)L'1)
\label{eq:6.KKmatrix} ,\\
M^{L'L}_{K\pm 2,K } &=&
 \textstyle\sqrt{{3\over 2}(\mu\mp K )(\mu \pm K+2)(1+\delta_{K,0})}\ 
(LK ,2, \pm 2|L',K\!\pm\!2), \nonumber
\eeqa 
It is then seen that 
 $\sqrt{2L+1} M^{L'L}_{K'K}$ is only equal to
$ (-1)^{L-L'} \sqrt{2L'+1} \big(M^{LL'}_{KK'}\big)^*$
when $L=L'$, as it should be for all $L$ and $L'$, for a unitary representation. 
Thus, it is profitable to initiate progression towards the construction of an orthonormal basis by a unitary transformation of the functions 
$\{\varphi^{(\lambda\mu)}_{KLM}\}$ to a new set 
\beq \Phi^{(\lambda\mu)}_{\alpha LM} := 
\sum_{K>0} \varphi^{(\lambda\mu)}_{KLM} U^{(L)}_{K\alpha} \eeq
such that the corresponding transformed matrices
\beq {\cal M}^{L'L}_{\beta\alpha} :=  \sum_{0\leq K,K'\leq \mu}  U^{(L')*}_{K' \beta } 
 M^{L'L}_{K'K}  U^{(L)}_{K\alpha } 
\eeq
are diagonal when $L'=L$, i.e., 
${\cal M}^{LL}_{\beta\alpha} = \delta_{\beta,\alpha}{\cal M}^{LL}_{\alpha\alpha}$.

We now claim that it remains only to make scale transformations of the wave functions, i.e.,
\beq \Phi^{(\lambda\mu)}_{\alpha LM} \to \Psi^{(\lambda\mu)}_{\alpha LM}
= k^{(L)}_\alpha \Phi^{(\lambda\mu)}_{\alpha LM} , \label{eq:5.knorms}
\eeq
to obtain an orthonormal basis for the Hilbert space of the VCS irrep.
This claim is substantiated by the observation that, in addition to reducing the subgroup chain ${\rm SU(3)} \supset {\rm SO(3)} \supset {\rm SO(2)}$,
the wave functions $\{\Phi^{(\lambda\mu)}_{\alpha LM}\}$ are also eigenfunctions of the Hermitian SO(3)-invariant operator
$\hat X := [ \hat L \otimes \hat {\cal Q}^{(2)} \otimes \hat L ]_0$, as the following will show.
In fact, as observed by Racah \cite{Racah62}, (to within norm factors) they are the unique simultaneous eigenfunctions of the  SO(3) and 
SO(2) Casimir invariants  and the SO(3)-invariant operator $\hat X$ and, as such, form an orthogonal basis for the finite-dimensional SU(3) irrep 
$(\lambda\mu)$.

According to the Wigner-Eckart theorem (given in any book on angular momentum theory, e.g. \cite{Rose57}), the coupled action of the
spherical tensor operator $\hat \Gamma({\cal Q}^{(2)}_\nu)$ on the wave functions of an orthonormal basis
$\{ \Psi^{(\lambda\mu)}_{\alpha LM}\}$, is expressed in terms of reduced matrix elements  by
 \beq [\hat \Gamma({\cal Q}^{(2)}) \otimes \Psi^{(\lambda\mu)}_{\alpha L}]_{L'M}
= \sum_\beta \Psi^{(\lambda\mu)}_{\beta L'M}
\frac{\langle (\lambda\mu){\beta L'} \| \hat{\cal Q}^{(2)}
\|(\lambda\mu) \alpha L\rangle}{\sqrt{2L'+1}} . \label{eq:6.WEa}
\eeq
The parallel equation for the operator $\hat X$, which is a coupled SO(3) tensor of angular momentum zero, is then
\beq \hat \Gamma(X)  \Psi^{(\lambda\mu)}_{\alpha LM}
= \sum_\beta \Psi^{(\lambda\mu}_{\beta LM}
\frac{\langle (\lambda\mu){\beta L} \| 
[ \hat L \otimes \hat {\cal Q}^{(2)} \otimes \hat L ]_0
\|(\lambda\mu) \alpha L\rangle}{\sqrt{2L+1}} . \label{eq:6.WEb}
\eeq
Now, the reduced matrix elements on the right side of this expression can be factored and determined to be proportional to the product of reduced matrix elements
\beq \langle L \| \hat L\| L\rangle \,
\langle(\lambda\mu) \alpha L \| {\cal Q}^{(2)}\|(\lambda\mu) \alpha L\rangle \,
\langle L \| \hat L\| L\rangle,
\eeq
with a proportionality factor that depends only on $L$.
It follows that \eq{eq:6.WEb} can be re-expressed 
\beqa \hat \Gamma(X)  \Psi^{(\lambda\mu)}_{\alpha LM}
&=& \sum_\beta \Psi^{(\lambda\mu)}_{\beta LM}\, f(L)\,
\langle (\lambda\mu){\beta L} \| \hat {\cal Q}^{(2)}
\|(\lambda\mu) \alpha L\rangle, \label{eq:6.WEc} \\
&=& 
\sqrt{2L+1}\, f(L)\,  
[\hat \Gamma({\cal Q}^{(2)}) \otimes \Psi^{(\lambda\mu)}_{\alpha L}]_{LM} ,
\eeqa
where  $f(L)$ is some function of  $L$.
This implies that $\Psi^{(\lambda\mu)}_{\alpha LM}$ is an eigenfunction of $\hat \Gamma(X)$ if and only if 
$[\hat \Gamma({\cal Q}^{(2)}) \otimes \Psi^{(\lambda\mu)}_{\alpha L}]_{LM}$ is proportional to $\Psi^{(\lambda\mu)}_{\alpha LM}$.
From this result, it follows that if we want $\Psi^{(\lambda\mu)}_{\alpha LM}$ to be proportional to $\Phi^{(\lambda\mu)}_{\alpha LM}$, we must similarly require that
\beqa \hat \Gamma(X)  \Phi^{(\lambda\mu)}_{\alpha LM} &=&  
\sqrt{2L+1}\, f(L)\,  
[\hat \Gamma({\cal Q}^{(2)}) \otimes \Phi^{(\lambda\mu)}_{\alpha L}]_{LM} 
\nonumber\\
&=& (2L+1)\, f(L)\,  
  {\cal M}^{LL}_{\alpha\alpha} \, \Phi^{(\lambda\mu)}_{\alpha LM}
\eeqa
which means that  $\Phi^{(\lambda\mu)}_{\alpha LM}$ is to be obtained by the unitary transformation of  ${M}^{LL}$ to a diagonal matrix
$ \sqrt{2L+1}\,{\cal M}^{LL}$.

To obtain an orthonormal basis for the irreducible Hilbert space 
$\mathcal{H}^{(\lambda\mu)}$ of VCS wave functions,
it now remains to determine the $k^{(L)}_\alpha$ norm factors appearing in 
\eq{eq:5.knorms},
with the understanding that any function $\Phi^{(\lambda\mu)}_{\alpha LM}$ that does not belong inside the space 
$\mathcal{H}^{(\lambda\mu)}$ of the \lfr{su}(3) irrep is to be assigned a zero norm factor.

To derive these norm factors, we make the substitution
$\Psi^{(\lambda\mu)}_{\alpha LM}
= k^{(L)}_\alpha \Phi^{(\lambda\mu)}_{\alpha LM}$
in the equation
\beq [\Gamma({\cal Q}^{(2)}) \otimes \Phi^{(\lambda\mu)}_{\alpha L}]_{L'M'}
= \sum_{\beta }\Phi^{(\lambda\mu)}_{\beta L'}\sqrt{2L'+1}\, \mathcal{M}^{L'L}_{\beta\alpha }
\eeq 
to obtain
\beq [\Gamma({\cal Q}^{(2)}) \otimes \Psi^{(\lambda\mu)}_{\alpha L}]_{L'M'}
= \sum_{\beta }\Psi^{(\lambda\mu)}_{\beta L'}\, 
\frac{k^{(L)}_\alpha}{k^{(L')}_\beta}\sqrt{2L'+1}\, \mathcal{M}^{L'L}_{\beta\alpha }.
\eeq 
Comparing with \eq{eq:6.WEa} then gives the identity
\beq \langle (\lambda\mu){\beta L'} \| \hat{\cal Q}^{(2)}
\|(\lambda\mu) \alpha L\rangle =(2L'+1)\mathcal{M}^{L'L}_{\beta\alpha }
\, \frac{k^{(L)}_\alpha}{k^{(L')}_\beta} .\eeq
For a unitary representation, these reduced matrix elements should satisfy the Hermiticity condition
\beq \langle (\lambda\mu){\beta L'} \| \hat{\cal Q}^{(2)}
\|(\lambda\mu) \alpha L\rangle = (-1)^{L-L'}
 \langle (\lambda\mu){\alpha L} \| \hat{\cal Q}^{(2)}
\| (\lambda\mu) \beta L'\rangle^* .
\eeq
Thus, for unitarity, the ratios of the norm factors are given by
\beq \left| \frac{k^{(L)}_\alpha} {k^{(L')}_\beta}\right|^2 = 
(-1)^{L-L'} \frac{2L+1}{2L'+1} \frac{\mathcal{M}_{\alpha\beta}^{LL'*}}
{\mathcal{M}_{\beta\alpha}^{L'L}}  
\eeq
and we obtain the explicit result
\beq \frac{\langle (\lambda\mu){\beta L'} \| \hat{\cal Q}^{(2)}
\|(\lambda\mu) \alpha L\rangle}{
\sqrt{(2L+1)(2L'+1)}} =\mathcal{M}^{L'L}_{\beta\alpha }
\, \sqrt{(-1)^{L-L'} \frac{\mathcal{M}_{\alpha\beta}^{LL'*}}
{\mathcal{M}_{\beta\alpha}^{L'L}} } .
\eeq
It will be noted that the only numerical calculation needed in the evaluation of this expression is the diagonalization of the $M^{LL}$  matrices, given explicitly by \eq{eq:6.KKmatrix}.

\section{The fundamentals of K-matrix theory} \label{sect:KmatTheory}

The construction of a reducible unitary irrep  of a Lie algebra \lfr{g} (or Lie group $G$) from a known finite-dimensional unitary irrep of a  subalgebra $\lfr{h}\subset \lfr{g}$ (or subgroup $H\subset G$) was achieved in the standard theory of induced representations \cite{Mackey68}.
In contrast, the VCS methods induce \emph{irreducible} unitary representations. 
This was enabled by the introduction of K-matrix theory 
\cite{Rowe84,Rowe95}, which determines the Hilbert space of the desired  irrep by the  construction of an orthonormal basis and the determination of its inner product..

The above examples have shown that renormalising an orthogonal set of  wave functions for a unitary irrep to obtain an orthonormal set is easy.
Thus, the primary task of   K-matrix theory is to determine an orthogonal basis for a VCS irrep.
As noted in Sect.\ \ref{sect:su3SO3},  any two eigenfunctions of a  Hermitian operator are necessarily orthogonal if they have different eigenvalues.  
Thus, a set of orthogonal wave functions is  derived if one has sufficient Hermitian operators to resolve any multiplicities.  
%
We now show that  K-matrix methods are simplified by the observation that  the product  $\hat S := \hat K \hat K^\dag$ is  a Hermitian operator.

\subsection{The S-matrix equations}

Let $\hat\Gamma$ denote a VCS representation of a Lie group $G$ and its
 Lie algebra $\lfr{g}$ that is  irreducible and unitary  with respect to an orthonormal basis for the Hilbert space  of VCS wave functions, $\mathcal{H}$.
In  proceeding to identity $\mathcal{H}$ and such a basis, we start with some larger space of wave functions, $\mathcal{F}$,  that is invariant under the action of $\hat \Gamma$ and  contains the space of VCS wave functions for the irrep 
$\hat\Gamma$ as an irreducible subspace.  
In practice there are natural ways to select the space $\mathcal{F}$, based simply on the requirement that it should be invariant under the action 
$\hat\Gamma$ of the Lie algebra  $\lfr{g}$,  as the 
examples considered in this review illustrate.  For example, for a scalar  irrep $\hat\Gamma$ of SU(3)  defined on a Hilbert space of functions of $\Omega\in {\rm SO}(3)$,   it would be appropriate to select $\mathcal{F}$ to be the space spanned by a basis for the regular representation of SO(3).
Thus, the concern of K-matrix theory is to identify the subspace, $\mathcal{H} \subset \mathcal{F}$ and the inner product for which it becomes the desired Hilbert space for the unitary VCS  irrep.

Let  $\{ \Psi_\nu\}$ denote an, as yet undetermined, orthonormal basis of VCS wave functions for $\mathcal{H}$ and let $\{ \varphi_n\}$ denote a convenient basis for $\mathcal{F}$.  The basis $\{ \varphi_n\}$ should have the property
that the  representation matrices, $\Gamma(X)$, defined by
\beq \hat\Gamma(X) \varphi_n = \sum_m \varphi_m \Gamma_{mn}(X), \quad
\forall\, X\in \lfr{g} ,
\eeq
are easily calculated.  This criterion is easily met if $\{ \varphi_n\}$ is an orthonormal basis for $\mathcal{F}$ with respect to some convenient Hermitian inner product.
The objective is then to determine $K$  matrices such that the desired orthonormal basis  functions for $\mathcal{H}$ have expansions
\beq \Psi_\alpha = \sum_n \varphi_n K_{n\alpha} ,\label{eq:4.Ktransform}\eeq
and facilitate  calculation of the  matrices $\gamma(X)$ of the unitary irrep on 
$\mathcal{H}$,  defined by
\beq
\hat\Gamma(X) \Psi_\alpha = \sum_\beta \Psi_\beta\, \gamma_{\beta\alpha}(X),
 \quad \forall\, X\in \lfr{g} .
\eeq
The matrix representation $\Gamma$, defined in terms of the convenient, but fundamentally arbitrary basis, $\{ \varphi_n\}$, for $\mathcal{F}$, is generally  neither unitary nor irreducible.
However, if we determine a $K$ matrix that maps the arbitrary basis for 
$\mathcal{F}$ to an orthonormal basis for $\mathcal{H}$ we also determine the  $\gamma(X)$ matrices.  

For a unitary irrep, the $\gamma$ matrices are required to satisfy the identity
$\gamma^\dag(X) = \gamma(X)$ for $X\in \lfr{g}$.%
\footnote{We follow the convention of quantum mechanics, 
     most commonly used in physics,  in which the elements of a 
     Lie algebra of observables, in a unitary representation, 
     are represented as Hermitian operators, e.g.,  position  
     and momenum observables of a particle are represented 
     by the Hermitian operators 
     $\{\hat x_j = x_j\}$ and $\{ \hat p = -{\rm i} \hbar \partial/\partial x_j\}$, 
     which satisfy commutation relations 
     $[\hat x_j, \hat p_k] = {\rm i} \hbar \delta_{j,k}$.}
It then follows  that
\beq \Gamma(X) K = K\gamma(X) \quad {\rm and}\quad
 K^\dag\Gamma^\dag(X)=
\gamma(X) K^\dag , \quad \forall \, X\in \lfr{g} .\eeq
It also follows that $K\gamma(X)K^\dag$ is equal to both
$\Gamma(X) KK^\dag$ and $K K^\dag\Gamma^\dag(X)$ and, hence,  that
\beq  S \Gamma^\dag (X)= \Gamma(X) S, \quad \forall\, X\in {\bf g},
\label{eq:4.Smatrixeq}\eeq
where $S:= KK^\dag$ is the matrix with elements
\beq S_{mn} = \sum_\alpha K_{m\alpha} K^*_{n\alpha} . \label{eq:S=KK*}\eeq
The objective is now to find systematic ways to solve \eq{eq:4.Smatrixeq} for the $S$ matrices.

\subsection{Making use of good quantum numbers} \label{sect:GQNos}

In solving \eq{eq:4.Smatrixeq}, it is advantageous to make use of the fact that, when the desired orthonormal basis wave functions for $\mathcal{H}$ reduce some specified chain of subgroups of $G$, they are  partially defined and labelled by the unitary irreps of the subgroups in this chain, which we assume to be known.  The subgroup labels then provide a set of 
  what we shall refer to as  
\emph{good quantum numbers}. 
   By this terminology we mean that functions labelled by different good quantum numbers are automatically orthogonal to one another.

Let $\kappa$ denote collectively a set of such good quantum numbers and  let $\alpha$ be the additional multiplicity label required to distinguish different wave functions with the same $\kappa$. Thus, we replace the label $\alpha$, as used above, by the double label $\kappa\alpha$ so that the orthonormal basis for
$\mathcal{H}$ is  a set $\{ \Psi_{\kappa\alpha}\}$.  Similarly, we can choose a basis for $\mathcal{F}$ by a set $\{ \varphi_{\kappa n}\}$.
Because functions in $\mathcal{F}$  with different values of the good quantum numbers of $\kappa$ are automatically orthogonal with respect to the inner product of $\mathcal{H}$, it follows that
  the $K$ and, hence, also the $S$ matrices, become block diagonal.  Thus, the basic K-matrix equations become
\beq \Psi_{\kappa\alpha }
= \sum_n \varphi_{\kappa n} K^{(\kappa)}_{n\alpha} ,
\label{eq:4.Ktransform2}\eeq
and
\beq  S^{(\kappa)} \Gamma_{\kappa,\kappa'}^\dag (X)= 
\Gamma_{\kappa,\kappa'}(X) S^{(\kappa')}, \quad \forall\, X\in {\bf g},
\label{eq:4.Smatrixeq2}\eeq
where $S^{(\kappa)}:= K^{(\kappa)}K^{(\kappa)\dag}$ and $\Gamma_{\kappa,\kappa'}(X)$ are, respectively, the submatrices with elements
\beq S^{(\kappa)}_{mn} 
= \sum_\alpha K^{(\kappa)}_{m\alpha} K^{(\kappa)*}_{n\alpha} , \quad
\Gamma_{\kappa m,\kappa' n}(X)
.\eeq
Equation (\ref{eq:4.Smatrixeq2}) is particularly useful because it gives recursion relations for the determination of the  matrices $S^{(\kappa)}$.
 Moreover, because these  matrices are Hermitian, they can be diagonalised by a unitary transformation of the 
 $\{\varphi^{(\kappa)}_n\}$ basis and brought to the form   
 $S^{(\kappa)}_{mn} = \delta_{m,n} \big(k^{(\kappa)}_n\big)^2$.
It  follows that a solution of the above equations for the $K^{(\kappa)}$ matrices are then given by $K^{(\kappa)}_{n\alpha}= \delta_{\alpha, n} k^{(\kappa)}_n$, where it is noted that because
$\mathcal{F}$ is generally bigger than $\mathcal{H}$ many of the 
 $k^{(\kappa)}_n$ are zero.
 
 \subsection{A more fundamental perspective on K-matrix theory}

The above K-matrix methods focus on determining an orthonormal basis of VCS wave functions.  The following approach gives an explicit expression for the above-defined  S matrices and an integral expression for the VCS inner product. 

Recall that a VCS wave function for a state $|\alpha\rangle$ in the Hilbert space, 
$\Hb$, for a given unitary irrep, $\hat T$, is defined by the overlap function 
\beq \Psi_\alpha(z) := \sum_\nu \xi_\nu \langle \nu|\hat T(z) |\alpha\rangle, 
\quad {\rm for}\, z\in Z, \label{eq:7.Psi}\eeq
where $\{ |\nu\rangle\}$ is an orthonormal set of basis vectors
 for an \emph{intrinsic} subspace $\Hb_0 \subset \Hb$, $\{ \xi_\nu\}$ are  wave functions for this set,  and  $Z\subset G^\Cb$ is chosen such that the Hilbert space, 
 $\Hb$, is spanned by the set of states
$\{ \hat T^\dag(z) |\nu\rangle, z\in Z\}$.%
\footnote{The intrinsic states can also be functionals 
     on a dense subspace of states in $\Hb$ .  The overlaps of 
     \eq{eq:7.Psi} are then well-defined for suitably chosen basis vectors, 
     $\{ |\alpha\rangle\}$ that span this dense subspace.}

Now the fact that the Hilbert space $\Hb$ is spanned by the states
$\{ \hat T^\dag(z) |\nu\rangle, z\in Z\}$, 
means that a complementary set of similar  wave functions, with vector values
$\{ \psi_\alpha (z) = \sum_\nu \xi_\nu \psi_{\nu\alpha}(z)\}$,
can be defined such that a vector $ |\alpha\rangle \in \Hb$ has expansion
\beq |\alpha\rangle = 
\sum_\nu\! \int_Z T^\dag(x) |\nu\rangle \xi^\dag_\nu \cdot \psi_\alpha(x)\, dv(x) , \label{eq:7.|psi>}\eeq
where $dv(z)$ is a convenient volume element for $Z$ and
$\xi^\dag_\mu \cdot\xi_\nu = \delta_{\mu,\nu}$.
Thus, the VCS wave function $\Psi_\alpha$ is  related to the function 
$\psi_\alpha$ by the equation
\beq  \Psi_\alpha (x)  := \hat S\psi_\alpha(x) = \int_Z S(x,y^*) \cdot \psi_\alpha (y)\, dv(y) ,\eeq
where $\hat S$ is the operator with kernel
\beq S(x,y^*) := \sum_{\mu\nu} 
\xi_\mu \langle \mu |\hat T(x) T^\dag(y) |\nu\rangle \xi^\dag_\nu . 
\label{eq:7.S(x,y)}\eeq
Moreover, an inner product for the Hilbert space $\Hb$ and a corresponding inner product for the space of $\{ \psi_\alpha\}$ functions is now given by
\beq \langle \alpha |\beta\rangle =
 \int_Z\!\int_Z \psi_\alpha^\dag(x) \cdot S(x,y^*) \cdot \psi_\beta(y)\, dv(x)\, dv(y) , 
 \label{eq:7.(psim,psin)}
 \eeq
Thus, the vectors $\{|\alpha\rangle\}$ generated by the functions 
 $\{\psi_\alpha\}$, in accordance with \eq{eq:7.|psi>}, form
 an orthonormal basis for the Hilbert space $\Hb$  if they 
 satisfy the orthogonalilty relationship
 \beq (\psi_\alpha , \hat S \psi_\beta ) :=
 \int_Z\!\int_Z \psi_\alpha^\dag(x) \cdot S(x,y^*) \cdot \psi_\beta(y)\, dv(x)\, dv(y) 
 = \delta_{\alpha, \beta},    \label{eq:7.(psim,o.n.)}
 \eeq
 We then obtain the notable result that the functions $\{ \psi_\alpha\}$ and the VCS wave function $\{ \Psi_\alpha = \hat S \psi_\alpha\}$ satisfy  the relationship
\beq (\psi_\alpha,\Psi_\beta) =
\int_Z \psi_\alpha^\dag(x) \cdot \Psi_\beta(y)\, dv(x) = \delta_{\alpha,\beta} .
\eeq
Thus, they are bi-orthogonal duals of each other relative to the inner product
$(\cdot ,\cdot)$.

We now consider the construction of an orthonormal basis of VCS wave functions $\{ \Psi_\alpha\}$ and 
their dual counterparts $\{ \psi_\alpha\}$.
First observe that inserting the identity operator 
$\hat I := \sum_\alpha |\alpha\rangle\langle \alpha |$ between the operators 
$\hat T(x)$ and $\hat T^\dag(y)$ in \eq{eq:7.S(x,y)}, reveals that
\beq  S(x,y^*)  = \sum_\alpha \Psi_\alpha(x) \Psi^\dag_\alpha(y).
\label{eq:S(x,y)}\eeq
Assuming we can derive the function $S(x,y^*)$,
 defined by \eq{eq:7.S(x,y)}, the determination of an orthonormal basis of VCS wave functions $\{ \Psi_\alpha\}$ from this expression is straightforward as follows.

Consider the $S$ matrix with elements
\beq S_{mn} := (\varphi_m, \hat S \varphi_n) . \label{eq:6.Smn} \eeq
It is Hermitian and, by \eq{eq:S(x,y)}, positive 
definite.
Thus, by a unitary transformation to a new basis
\beq \Phi_\alpha =  \sum_n  \varphi_n U_{n\alpha} , \eeq
it can be brought to the diagonal form
\beq S_{\alpha\beta} = \sum_{mn} U^*_{m\alpha} S_{mn} U_{n\beta} 
= \delta_{\alpha,\beta} k_\alpha^2 .\eeq

The VCS wave functions $\{ \Psi_\alpha\}$  and their $\{ \psi_\alpha \}$ counterparts are then defined, for the non-zero values of $k_\alpha$, by
\beq \Psi_\alpha := k_\alpha \Phi_\alpha , \quad 
\psi_\alpha = \frac{1}{k_\alpha}  \Phi_\alpha .\eeq
Finally, matrix elements of the VCS representation in an orthonormal basis  are determined from 
\beq  \langle \alpha |\hat T(g) |\beta\rangle = 
(\psi_\alpha, \Gamma(g) \Psi_\beta) 
= \frac{k_\beta}{k_\alpha} (\Phi_\alpha , \hat\Gamma(g) \Phi_\beta) , \quad
g\in G .\eeq

The above algorithm simplifies considerably when there are good quantum numbers (as defined in Sect.\ \ref{sect:GQNos}).   
If $\alpha$ is replaced by a double index $\kappa\alpha$, where 
$\kappa$ denotes  collectively a set of good quantum, then $S$ is expressible as the sum
\beq S(x,y^*) = \sum_{\kappa\alpha} \Psi_{\kappa\alpha}(x) 
\Psi^\dag_{\kappa\alpha}(y) .\eeq
Then, with an expansion of an orthonormal basis of VCS wave functions
\beq \Psi_{\kappa\alpha}(x) = \sum_n \varphi_{\kappa n}(x)K^{(\kappa)}_{n\alpha} ,\eeq
in a basis, $\{ \varphi_{\kappa\alpha}\}$, for $\mathcal{F}$ that is orthonormal with respect to the inner product
\beq ( \varphi_{\kappa m},\varphi_{\kappa n}) := 
\int_Z \varphi^\dag_{\kappa m}(z) \cdot \varphi_{\kappa' m}(z) \, dv(z)
= \delta_{\kappa',\kappa} \delta_{m,n} ,\eeq
we obtain $S$ as a sum  $S = \sum_\kappa S^{(\kappa)}$ with
\beqa S^{(\kappa)} (x,y^*) 
&=& \sum_{mn\alpha} \varphi_{\kappa m}(x) K^{(\kappa)}_{m\alpha}
K^{(\kappa)*}_{n\alpha} \varphi^\dag_{\kappa n}(y) \nonumber\\
&=& \sum_{mn} \varphi_{\kappa m}(x)
S^{(\kappa)}_{mn}  \varphi^\dag_{\kappa n}(y) .\eeqa
Thus, a unitary  transformation that brings the submatrices $S^{(\kappa)}$ to the diagonal form 
\beq S^{(\kappa)}_{\alpha\beta} 
= \delta_{\alpha,\beta} \big(k^{(\kappa)}_\alpha\big)^2 ,\eeq
defines the orthonormal wave functions for the non-zero values of 
$k^{(\kappa)}_\alpha$
\beq \Psi_{\kappa\alpha} := k^{(\kappa)}_\alpha \Phi_{\kappa\alpha} , \quad 
\psi_{\kappa\alpha} = \frac{1}{k^{(\kappa)}_\alpha} \Phi_{\kappa\alpha} , \quad
{\rm with} \quad \Phi_{\kappa\alpha} = 
\sum_n \varphi_{\kappa n} U^{(\kappa)}_{n\alpha} ,\eeq
and we can proceed as above.

The usefulness of the above are illustrated by their application to
scalar coherent state representations \cite{RoweR02}.  For example, for the holomorphic SU(1,1) irreps considered in Sect.\ \ref{sect:su11}, it is determined that
\beq S(x,y^*) 
= \langle \lambda 0| e^{x\hat S_-} e^{y^* \hat S_+} |\lambda 0\rangle
= (1-xy^*)^{-\lambda}, \eeq
which, for $|x|$ and $|y| <1$, has the Taylor expansion
\beq S(x,y^*)  = \sum_\nu  \frac{(\lambda + \nu -1)!}{(\lambda-1)! \nu!} 
\big(xy^*\big)^\nu = \sum_\nu \Psi_\nu (x) \Psi^*_\nu(y),
\eeq
and gives
\beq \Psi_\nu (z) = \sqrt{\frac{(\lambda +\nu -1)!}{(\lambda-1)! \nu!}} \, x^\nu, \quad \nu = 0,1,2, \dots \,;\eeq
cf.\ \eq{eq:2.su11wfn}.
Thus, an orthonormal basis of scalar coherent state wave functions for the SU(1,1) irrep with lowest weight $\lambda$ is given by the set
$\{ \Psi_\nu , \nu=0,1,2, \dots\}$.

\section{Concluding remarks} \label{sect:conc}

This article has given  representative examples of a few uses of scalar and vector coherent state representations of Lie algebras.  A much greater diversity of representations could have been given.  For example, the standard 
Schr\"odinger representation of the Hilbert space $\mathcal{L}^2(\Rb^3)$ for a particle moving in the Euclidean space $\Rb^3$  can be seen as a coherent state representation \cite{Rowe96} with wave functions expressed, for all vectors 
$|\psi\rangle$ in the dense subspace of continuously differentiable functions in 
$\mathcal{L}^2(\Rb^3)$, by
\beq \psi({\bf r}) := \langle 0| 
\exp \big( - \frac{\rm i}{\hbar} {\bf r} \cdot \hat{\bf p}) |\psi\rangle ,\eeq
where $\hat p = -{\rm i} \hbar \nabla$ and $\langle 0|$ is the Dirac delta functional
defined by 
\beq \langle 0 |\psi\rangle = \int \delta({\bf r})\, \psi({\bf r}) \, d{\bf r} = \psi(0).\eeq
The Sch\"odinger representation of a particle with intrinsic spin can likewise be seen as a VCS representation \cite{Rowe96}.
The use of such functionals to define dense subspaces of scalar and VCS wave functions can also be used profitably to construct representations other than those of a discrete series (i.e., other than those that appear in the regular representation).  Example of such representations occur widely in physics for systems whose dynamical groups are semi-direct products with Abelian normal subgroups, e.g., Euclidean groups, space groups, the Poincar\'e group, and rotor model groups. 

The examples considered here have  been restricted to the unitary irreps of Lie algebras.  However, they have natural extensions to non-unitary representations of Lie groups and their Lie algebras, such as carried by the finite tensor operator representations of non-compact Lie algebras.  They also have natural extensions to super-algebras \cite{LeBlancR89,LeBlancR90}.

Another application is to the calculation of the Clebsch-Gordan coefficients needed to derive the decomposition of a tensor product of two representations of a Lie group into a sum of irreps \cite{RoweB00,BahriRD04}.  
These coefficients appear in the construction of  the irreps of the direct product of two copies of a group, $G\times G$. 
 If we denote an element of this product by a pair $(g_1,g_2)$.
 then the required Clebsch-Gordan coefficients are obtained by the construction of the irreps of $G\times G$ in a basis that reduces the subgroup 
$\tilde G  := \{ (g,g), g\in G\}$ isomorphic to $G$.

Yet another application is to obtain accurate contraction limits to Lie groups and their Lie algebras.  Such contraction limits occur when some parameter in the definition of a Lie group or its representation goes to zero or to infinity.  This happens, for example, for large-dimensional representations and for large values of some component of a highest or lowest weight.
Contractions of this kind are important in physics because they often lead  to classical insights and to simple but highly accurate approximations to an otherwise complex system.
Familiar contraction limits occur in non-relativisitic limits and, in quantum mechanics, when the scales of interest are large compared to those imposed by the uncertainty principle.
Contraction limits are realised, for example, when the low-energy states of a system behave as though the spectrum generating algebra for the system were a simple Heisenberg or boson algebra, as in a normal-mode theory of small amplitude vibrations.  Other contraction limits are realised when the low-energy states of a system behave like those of a rotor.
For example, holomorphic representation in which a Lie algebra of observables is expressed in terms of a set of complex variables $\{ z_i\}$ and their derivatives 
$\{ \partial/\partial z_i\}$, with commutation relations
\beq [\partial/\partial z_i\ , z_j ] = \delta_{i,j} .\eeq
can clearly be expressed as boson expansions by the substitution
 $z_i \to c^\dag_i$ and $\partial/\partial z_i \to c_i$ wtih 
 $[c_i, c^\dag_j] = \delta_{i,j}$.
 The construction given above for the VCS irreps of SU(3) in a basis of rotor model wave functions, likewise enables a contraction of SU(3) to a semi-direct product rotor model group.

These many applications demonstrates the  power of VCS representation theory as a tool in the application of symmetry methods in physics. It would therefore be surprising if it were not also useful in mathematics, at least as a unifying theory that naturally incorporates the theories of induced representations and geometric quantisation.

\section*{Acknowledgements}
The author is pleased to acknowledge  valuable discussions and collaborations on this subject with J. Repka.

\bigskip
\section*{References}
\bibliography{CMQMvcs}

\end{document}